\documentclass[journal]{IEEEtran}

\usepackage[utf8]{inputenc}
\usepackage{cite}
\usepackage[mathscr]{euscript}
\usepackage{amssymb}

\usepackage{amsmath}
\usepackage{breqn}

\usepackage{setspace}
\usepackage{epsf}\epsfverbosetrue
\usepackage{graphics,epsfig}
\usepackage{graphicx,epsfig}
\usepackage{epsfig}
\usepackage{multirow}
\usepackage{alltt}
\usepackage{subfigure}
\usepackage{tcolorbox}
\usepackage{float}
\usepackage{url}
\usepackage{graphicx}
\usepackage{verbatim} 
\usepackage{footnote} 
\usepackage{latexsym} 
\usepackage{sidecap} 
\usepackage{wrapfig}
\usepackage{eso-pic}
\usepackage{fix-cm}
\usepackage[mathscr]{euscript}
\usepackage{algorithm}
\usepackage{algorithmic}
\usepackage{color}
\usepackage{wrapfig}
\usepackage{multicol}

\usepackage{amsfonts}
\usepackage{amssymb}
\usepackage{amsmath}

\newtheorem{theorem}{\textbf{Theorem}}

\newtheorem{Lemma}{\textbf{Lemma}}

\usepackage{layout}

\usepackage{amsmath}
\usepackage[
  locale = DE 
]{siunitx}

\usepackage{textcomp}
\usepackage{multirow}
\usepackage{mathtools}
\usepackage{soul} 
\usepackage{amsmath} 
\usepackage{verbatim}
\usepackage{csvsimple} 
\usepackage{array}
\usepackage{subfig}
\usepackage[normalem]{ulem}
\usepackage{booktabs}
\newcolumntype{P}[1]{>{\RaggedRight\arraybackslash}p{#1}}

\usepackage{pifont}

\newcommand{\snp}[1]{\textcolor{black}{#1}}

\usepackage{amsmath}

\begin{document}

\title{\snp{VPT: Privacy Preserving Energy Trading and Block Mining Mechanism for Blockchain based Virtual Power Plants}}

\author{Muneeb Ul Hassan, Mubashir Husain Rehmani, and Jinjun Chen
\thanks{M. Ul Hassan and J. Chen are with the Swinburne University of Technology, Hawthorn VIC 3122, Australia  (e-mail:  muneebmh1@gmail.com; jinjun.chen@gmail.com).}
\thanks{M.H. Rehmani is with the Munster Technological University (MTU), Ireland (e-mail: mshrehmani@gmail.com).}
\thanks{Please direct correspondence to M. Ul Hassan.}
}



\maketitle

\begin{abstract}

The desire to overcome reliability issues of distributed energy resources (DERs) lead researchers to development of a novel concept named as virtual power plant (VPP). VPPs are supposed to carry out intelligent, secure, and smart energy trading among prosumers, buyers, and generating stations along with providing efficient energy management. Therefore, integrating blockchain in a decentralized VPP network emerged as a new paradigm, and recent experiments over this integration have shown fruitful results. However, this decentralization also suffers with energy management, trust, reliability, and efficiency issues due to the dynamic nature of DERs. In order to overcome this, \snp{in this paper, we first work over providing an efficient energy management strategy for VPP to enhance demand response, then we propose an energy oriented trading and block mining protocol and name it as proof of energy market (PoEM). To enhance it further, we integrate differential privacy in PoEM and propose a \textbf{P}rivate PoEM (PPoEM) model. Collectively, we propose a private decentralized VPP trading model} and named it as \textbf{Vi}rtual \textbf{P}rivate \textbf{T}rading (VPT) model. We further carry out extensive theoretical analysis and derive step-by-step valuations for market race probability, market stability probability, energy trading expectation, winning state probability, and prospective leading time profit values. Afterwards, we carry out simulation-based experiments of our proposed model. The performance evaluation and theoretical analysis of our VPT model make it one of the most viable models for blockchain based VPP networks as compared to other state-of-the-art works.
\end{abstract}


\section{Introduction}

In order to efficiently manage the growing number of distributed energy resources (DERs) and keep their management separate from the main grid, researchers introduced a novel concept of virtual power plant (VPP). VPP can be defined as an entity which integrates multiple DERs in order to control them in a uniform manner. VPPs are designed to carry out various tasks ranging from load monitoring, load control, peak management, \snp{energy trading, demand side management, etc~\cite{spref09}. These VPPs efficiently support the integration of different variable DERs into energy markets such} as solar photovoltaic panels, electric vehicles (EVs), controllable loads, storage batteries, etc. DERs participate in the energy markets in presence of multiple VPPs and carry out joint energy trading. VPPs are responsible to carry out energy trading of DERs from the prosumers to grid stations. Therefore, they are developed and programmed in such a way that they maximize revenue and enhance controllability factor in order to manage everything optimally~\cite{spref02}.
~The VPP trading model suffers from three major drawbacks in the form of incentive compatibility, trust, and privacy. Firstly, every trader wants to enhance their revenues to maximum, similarly, everyone wants to know the details of \snp{their trading in this modern world and want to ensure that they} are getting maximum benefit from their business. \snp{Therefore, the security and trust play an important factor in order to attract} more buyers towards some specific application~\cite{relnew06}. In order to enhance trust and security in the network, blockchain technology came up as a rescuer~\cite{spref16}. \snp{The decentralized and immutable nature of blockchain ensures that every buyer is being treated equally and no one is being given} unnecessary favour. Blockchain in this scenario ensures that every user has control of their data, and they can verify their transactions anytime without having any type of risk of cheating. \\ 
Furthermore, the consensus mechanism in blockchain plays an important role in validating and approving the transaction because every transaction needs to pass through verified miners in order to get added inside the block~\cite{spref03}. Similarly, mining in blockchain also ensures that all the blocks being recorded in a ledger are legit and does not contain any anomalous transaction/information~\cite{spref05}. Traditional \snp{blockchain networks work over proof-of-work (PoW) mining based consensus to determine winning miner which is not suitable for VPP based models due its computational} complexity~\cite{spref04}. Moreover, since yet, \snp{a blockchain mining mechanism that is purely developed from the perspective of energy trading of decentralized VPPs has} not been discussed previously. Therefore, in this article, \snp{we develop a distributed consensus miner determination mechanism purely oriented towards energy trading of VPPs and named it as Proof of Energy Market (PoEM).}\\ 
To add it further, we overcome the issues of \snp{privacy of blockchain by integrating differential privacy in PoEM mechanism and proposed a Private Proof of Energy Market (PPoEM) mechanism in which the privacy of buyers, sellers, and VPPs will be protected using the concept of differential privacy perturbation.} Furthermore, to incentivize all participating prosumers and buyers along with management of demand response, we propose a VPP monitoring based energy trading model that motivates DERs to sell maximum energy during a system when energy demand is high. \snp{Collectively, we propose a complete blockchain based VPP trading model and named it as virtual private trading (VPT) model. }

\vspace{-1mm}
\subsection{Related Works}

DERs is a well-researched domain and plenty of research has been carried out to efficiently manage operations of DERs especially focusing on energy trading and management. For example, authors in~\cite{spref06} work over clustering formation of heterogeneous on power requirements of VPP smart grid scenario. Similarly, another work targeting risk constrained management of energy for enhancement of demand response via VPP has been carried out by researchers in~\cite{spref07}. Another work that targets distributed dispatch of VPPs under cyber threats has been carried out by researchers in~\cite{relref06}. Nowadays, another novel shift in paradigm has happened and researchers are integrating decentralized blockchain technology with VPPs. In this scenario, a  detailed work has been carried out by authors in~\cite{relnew01}, which discussed possibilities and future trends of this integration. However, \snp{two major issues of effective energy oriented miner determination for consensus and privacy preservation in blockchain still needs to be addressed. To the best of our knowledge, our work is the first pioneering work towards integration of a novel and private block mining mechanism from the perspective of decentralized energy trading via VPPs.} For more details regarding privacy issues and integration of differential privacy in blockchain and other scenarios, we recommend readers to study~\cite{prelref04}.


\subsection{Key Contributions}
The key contributions of our work are as follows:

\begin{itemize}

\item We work over integration of blockchain in VPP scenario, and developed a \snp{complete three layered VPP model operating over permissioned blockchain}
\item \snp{We propose a VPP system state determination model for efficient energy trading and price determination by VPPs.}
\item We propose~\snp{PoEM and PPoEM mechanisms via which participants (such as buyers, sellers, and VPPs) can easily trade electricity without the} risk of losing or compromising their private data along with enhancing trust in the network.
\item We also enhance social welfare for both \snp{buyers and sellers along with incentivizing VPPs for energy trading tasks in a private manner.}

\end{itemize}

The remainder of the paper is organized as follows: Section 2 discusses system model and \snp{functioning of VPT Model. Section 3 provides discussion about development of PoEM and PPoEM algorithms. Furthermore, Section 4 gives extensive theoretical analysis for differential privacy, security, market and profit related probabilities. Afterwards, experimental simulation results, analysis,} and behaviour of PoEM and PPoEM is given in section 5. Finally, section 6 provides the conclusion of the paper.

\begin{figure*}[ht]        
\centering
\includegraphics[scale = 0.1]{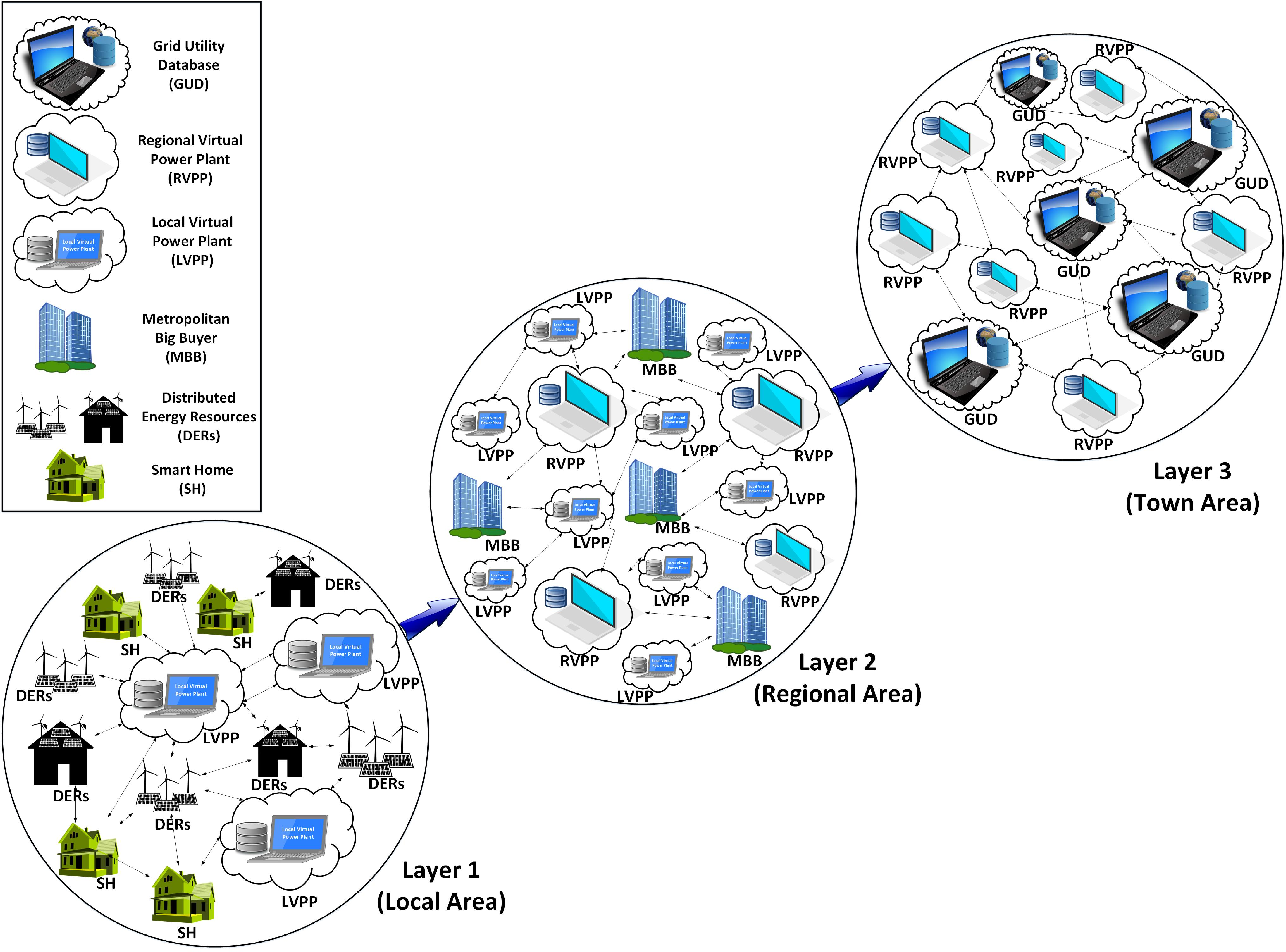}
 \caption{Blockchain based Virtual Power Plant Scenario for Incentive Compatible Energy Trading.}      
  \label{fig:fig01}   
\end{figure*}

\section{System Model and Functioning} \label{KeySection}

VPPs are the most critical participant in our proposed VPT strategy, and all decentralized energy trading functionalities revolve around the efficient functioning of VPPs. In order to provide a complete picture of the VPT scenario to our \snp{readers, in this section, we discuss preliminaries such as motivation, problem statement, system model \& structure and adversary model in detail.}

\subsection{Functioning of Virtual Power Plants}
The notion of VPP was introduced to stimulate a platform in which DERs can be managed efficiently without involvement of traditional centralized grid~\cite{relref07}. The objective of VPPs based smart grid is to develop such an environment in which DERs will be given more decision flexibility along with enhancement of demand response of that specific area by implementing specific policies~\cite{spref08}. VPPs are able to monitor, control, forecast, dispatch, and optimize the consumption and generation of DERs in the specified region. To discuss it further, in our scenario \snp{work over the specific aspect of energy trading via VPPs. As discussed above, VPPs will be able to develop policies, along with optimization of  demand response. Therefore, in our scenarios VPPs will manage energy trading by carrying out a double auction between buyers (homes \& buildings) and sellers (DERs). In this trading, VPPs can make policies in which they can incentivize prosumers in order to motivate them to sell electricity at times of high demand hours in order to enhance demand side management.}

\subsection{Differential Privacy}
The concept of “Differential Privacy” as a privacy preservation strategy first came into discussion after its successful implementation in statistical databases in 2006 by C. Dwork~\cite{intref00}. Differential privacy was developed to ensure that any query evaluator will not be able to get exact information of a specific individual within a dataset~\cite{spref24}. According to an informal definition, differential privacy claims that addition, modification, or deletion of a single individual record does not have any significant effect over the result of any query analysis~\cite{spref10}. In our VPT model, we \snp{use both Laplace and Exponential mechanism of differential privacy} to privatize the mining and auction process, which is discussed in next sections.

\subsection{Motivation of our Work} 

The motivation of our VPT strategy and novel mining mechanisms is as follows:
\begin{itemize}
\item Traditional \snp{energy auctions do not incentivize prosumers} if they sell energy during peak demand hours~\cite{spref11}. We develop an incentivizing mechanism which will provide benefits to energy traders if they sell during peak hours.
\item Typical VPP based energy trading mechanisms do not incorporate blockchain in their proposed system model~\cite{spref12}. However, in our VPT strategy, we use permissioned blockchain to enhance trust.
\item Conventional mining mechanisms used in VPP based energy trading are \snp{not incentivising VPPs on the basis of energy they are trading~\cite{relnew01, spref04}. Mining phenomenon in our PoEM consensus mechanism motivates VPPs to carry out maximum energy trading by choosing miners on the basis of energy they are trading.}
\item \snp{Traditional block mining and trading mechanisms of VPPs does not incorporate privacy preservation from the perspective of both; buyers and VPPs. Our proposed PPoEM mining mechanism uses advantages of differential privacy and ensures privacy preservation of VPPs and buyers.}
\end{itemize}

\subsection{System Model \& Structure}\label{SysMod}

We divide the complete system model of VPT into three layers that target and cover a complete VPP based energy network (given in Fig.~\ref{fig:fig01}). Starting from the local area energy network, each DER is connected to each VPP in the prescribed area, e.g., this prescribed area could be a suburb, or combination of few suburbs (depending upon the density). DERs can provide their available energy to the local VPP (LVPP) of their choice for auction, and they will do so depending upon the incentives and rates each VPP is giving. For example, `LVPP 1’ charges \$5 per transaction and `LVPP 2’ charges \$3 per transaction, then definitely DERs will tend to go for the one charging less transaction fee (Here \$ is only used to provide a generalist point of view, although we are using VPP coin in our VPT model). However, on the other hand `LVPP 1’ can provide some other incentives, etc, in order to attract maximum customers. Similarly, these VPPs can also provide incentives to the buyers and can attract more buyers than others.\\ Similarly, from the figure, it can also be observed that each smart home is also connected to all LVPPs of the specified area, which means that they can purchase energy from the LVPP of their choice.  All the local regions will be categorized by keeping in view the perspective of energy internet (EI), a detailed discussion about EI can be found in the work carried out by Want~\textit{et al.} in~\cite{keyref01}. 
Similarly, in layer 2 of our proposed VPT trading model, LVPPs are connected with MBBs and regional VPPs (RVPP). Here, metropolitan big buyers (MBBs) can request RVPPs if they require a large amount of energy for a specific time-slot (e.g., in case of a specific event, etc.). This task can further be distributed to multiple LVPPs by RVPPs in order to meet the demand. Here the competition is among RVPPs, and each RVPP can provide incentives to attract as many MBBs as they can. In this layer, RVPPs will be mining nodes and competition will be among them. Moving further to the layer 3 of our VPT energy trading model, in which these RVPPs are connected with grid utility databases (GUD), and RVPPs can trade their energy with GUDs as well. Here GUDs will  incentivize these RVPPs in order to generate maximum profit after selling this energy to their consumers, similarly, in this layer, GUDs will be the mining nodes. 



\subsection{\snp{Adversary Model}}
\snp{In our proposed VPT model, bidders and sellers submit their truthful bids and asking prices to VPPs in order to maximize their social welfare. Similarly, at the time of mining, VPPs report their truthful energy trading details to the mining authority to enhance trust in the blockchain network. However, if this truthfulness is not maintained due to some adversarial impact, then the level of trust will decrease in the network which will have a direct impact on the functionality. In our VPT model, we divide adversaries into two types, one is from perspective of adversarial objectives during bidding and auction, while second is from perspective of adversarial objectives during mining process. }
\subsubsection{\snp{Adversarial Objectives During Double-Sided Auction}}
\snp{From the perspective of adversarial objective during auction, the major information that adversary during auction process is aiming to infer is the private bids of buyers. Alongside private bids, the adversary is also interested in getting the private asking price and generation values of energy sellers as well. Leaking these private values not only has an effect on the personal privacy of sellers and buyers but it will also have a direct impact on the fairness and privacy of the auction market. The first reason behind this impact is that when an adversary will have fine grained data about asks, generation, bids, etc of an auction market, then it is easy for the adversary to infer and get more personal information of a particular household by learning from the data. For instance, just by carrying out learning and comparison via inference attack on valuations and usages, an adversary can infer more private data such as generation per hour, living habits, environmental factors,  trading agenda, environmental factors, etc~\cite{spref17}. Secondly, the adversary is also considered curious about the processes involved in auction in order to play strategically. E.g., getting accurate information about valuations, asks, availability, and other similar values which are used by VPPs to determine hammer/threshold price of the auction. This inference of threshold price can be of strategic advantage to an adversary, because an adversary can then participate in auction to gain high profit, which in turn will reduce profit of other participants.}

\begin{figure}{ht}
\centering
\includegraphics[scale=0.6]{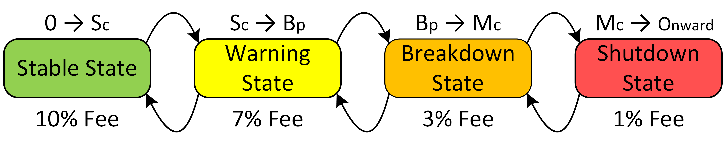}
\caption{Transition Between System State of Local VPP}
\label{SysState02}
\end{figure}

\subsubsection{\snp{Adversarial Objectives During Blockchain Mining}}
\snp{From perspective of mining in blockchain mining, it is important to understand that majority of the time, VPPs do not want other VPPs to know their exact trading information because of the competition among them. In an energy oriented mining mechanism, the major objective of an adversary is to find out the highest trading VPP in order to study the trading strategies of that VPP. An adversarial node can try to infer privacy of high trading VPP in order to get deeper insights about the type of advertising, marketing, and trading strategies the specific VPP is opting out. So, every VPP (especially if it is high trading VPP) wants to keep these strategies private. Therefore, if one gets to know that a specific VPP won the mining election just because of the reason that he was among some high trading VPPs, then the adversarial competitors will focus to infer into that specific VPP because this inference can be of strategic advantage to adversary. Contrarily, if it is not clear and there are chances that a VPP with low trading score can also become winner due to differential privacy, then this adversarial risk reduces to minimum. }

\subsection{Motivation to Use Permissioned Blockchain}
\snp{Our VPT model works over the phenomenon of permissioned decentralized blockchain technology. The motivation to use permissioned blockchain instead of a traditional database arises due to the need for trust in the network. Because contrary to traditional distributed databases, our permissioned VPT blockchain networks enhance trust by providing an append-only copy of decentralized ledger to all its nodes. The ledger is an append-only structure and data inside cannot be changed once it gets stored because even VPPs do not have the permission to modify this data. VPPs can only append a new block in the chain but cannot change the previous content of blocks which was possible in traditional distributed databases.}\\
\snp{Moreover, the reason behind using permissioned blockchain instead of a permissionless blockchain is because energy and smart grid is a sensitive domain and it is important to control who can join the network. In our VPT model, only the designated nodes which have a smart meter and are capable of trading/purchasing energy can join the network. In order to carry out all these permissioned operations, a permissioned blockchain is required instead of a permissionless blockchain. Furthermore, the permissioned nature is also used to control malicious behavior of nodes as well, such as over provisioning of energy, etc. Detailed security analysis of our work is presented in Section~\ref{SecAnalysis}.}

\subsection{Ledger \& Block Model} \label{ledger}

\snp{A well-maintained decentralized distributed ledger is considered to be one of the most important aspects of blockchain. One of the most critical things in ledger formation and maintenance is its phenomenon of recording a transaction. As discussed in the section above, our VPT model contains three different types of ledgers at each layer. These ledgers are controlled by different authoritative nodes. For example, in the first layer of Fig.~\ref{fig:fig01}, the control of the ledger is in the hand of LVPPs, and it can only be visualized/seen by participating smart homes and DERs. Similarly, if we have one more local area energy network, its ledger will be controlled by its own LVPPs, and the LVPPs of the first network will not have any control over the other area. Moving further to the next layer, RVPPs have a separate ledger model with LVPPs, in which RVPPs serve as controlling nodes, and have the right to mine the transactions. However, LVPPs are only readers in this second layer and do not have mining rights. These RVPPs can also read the data from the first local energy network layer, however they cannot modify or mine within the first layer and require approval from authorities in order to do so. Here, authorities are grid utility databases which are managing the complete system. The next layer is the third and final layer of our VPT strategy, in this layer all RVPPs are connected with the main grid utility and they have a separate ledger for their own transactions. Basic DERs, smart homes, and LVPPs do not have access to this ledger, and only  RVPPs and GUDs can view the content and participate in mining of this ledger.\\}
\snp{Moreover, the block structure in our ledger comprises energy transactions from trading nodes. Each transaction of our VPT model contains information about the microgrid seller, smart home or MBB buyer, amount of energy being purchased, amount of VPP coin being spent, and time slot.}

\begin{algorithm}[t]\scriptsize
\caption{System State Determination by VPP}
\label{AlgoSysState}

\textbf{Input:}  $G_E, S_l, B_p, M_c$ \\
\textbf{Output:} $S_s$ \\
\textbf{Call:} EnergyDetection($G_E, S_l, B_p, M_c$)\\

\textbf{FUNCTION} $\gets$ EnergyDetection($G_E, S_l, B_p, M_c$)
\begin{algorithmic}[1]
\STATE $G_E \gets$ Real-Time Demand From Grid Utility
\IF {($0 \leq G_E \leq S_l$)}
\STATE RT$_{tx}$ =  RT$_m$ = 10\%
\STATE $S_s \gets$ Stable State
\ELSIF {($S_l < G_E \leq B_p$)}
\STATE RT$_{tx}$ =  RT$_m$ = 7\%
\STATE $S_s \gets$ Warning State
\ELSIF {($B_p < G_E \leq M_c$)}
\STATE RT$_{tx}$ =  RT$_m$ = 3\%
\STATE $S_s \gets$ Breakdown State
\ELSIF {($M_c < G_E $)}
\STATE RT$_{tx}$ =  RT$_m$ = 1\%
\STATE $S_s \gets$ Shutdown State
\ENDIF

\STATE \textbf{return} $S_s$

\end{algorithmic}
\end{algorithm}


\section{Development of VPT Energy Trading and Consensus Mining Model} \label{consensus}
In this section, \snp{first we discuss our VPT model from the perspective of system state determination and then we discuss the development and functioning of our PoEM and PPoEM trading and consensus miner determination algorithms.} 

\subsection{System State Determination by VPPs}
To incentivize selling \snp{prosumers, VPPs can make policies} that encourage more prosumers to sell their stored energy at the time of high demand~\cite{spref19}. In order to do so, we work over integration of features of system state determination and incentivization at the level of LVPPs. A detailed formulation of state determination has been provided in Algorithm~\ref{AlgoSysState} and Fig.~\ref{SysState02}. In the given algorithm, firstly, the values of stable state, breakdown state, and shutdown state are fed to LVPP. Afterwards, VPP regularly monitors the grid energy that is being used in the specified area. If the energy usage is under a stable region, VPPs charge a 10\% fee for mining and transaction reward. However, in case of need when the system is in warning, breakdown, or shutdown state, VPP can reduce the fee to 7\%, 3\%, and 1\% respectively to encourage maximum microgrid prosumers to sell their energy. Extensive evaluation of this approach on real datasets have been provided in Section~\ref{perfsection}.

\begin{algorithm}[t]\scriptsize
\caption{Proof of Energy Market (PoEM) Algorithm}
\label{algoDP}

\textbf{Input:}  b, m, N, E, a, MR, VPP, RT$_{tx}$, RT$_m$

\textbf{Output:} W$_{VPP}$, SW${_b}$, SW${_s}$, W${_b}$, W${_s}$

\vskip 2mm
\hspace*{4mm} (1) \underline{Carrying out Double Auction}
\vskip 1mm

\begin{algorithmic}[1]

\STATE \textbf{max$_{bid_{(s)}} \gets argmax[sort(b)]$}
\FOR{\textbf{each}~\textit{seller} j $\gets$ 1 to $S_{max}$}

\FOR {\textbf{each}~\textit{buyer} k $\gets$ 1 to $N_{max}$}

\IF {(b $\geq$ a \& b == max$_{bid}(j)$)}
\STATE \textbf{Calculate} $j^{th}$ energy slot winner $({W}_j)$ w.r.t rule of allocation
\STATE $W_j(x) = argmax_b \sum_{k \in N}{X_k(b)}$\\
//~$W_j(x)$ is the selected winner for slot $E(j)$
\ELSE 
\STATE \textbf{return} 'bid did not match the ask'
\STATE \textbf{break;}
\ENDIF

\STATE \textbf{Calculate} Price (F$_p$) of $k^{th}$ buyer w.r.t payment rule
$I_p$ = $b(k)$
\ENDFOR
\STATE~\textbf{Append} winner ID, price, energy slot
\STATE \textbf{Append} $W_s[I_d, {F_p}_i,  S_{id}]$

\ENDFOR

\vskip 1.2mm
(2)~\underline{Compute Social Welfare, Transaction Fee \& Mining Fee}
\vskip 1mm
\STATE RT$_{tx}$, RT$_{m} \gets$ via $S_s$ from Algorithm 1
\FOR {$j \gets$ 1 to $W_{s(max)}$}
\STATE \textbf{Compute} Transaction Fee via  RT$_{tx}$ 
\STATE T$_{x_{f}}$ = F$_p(j) * ~$RT$_{tx}$

\STATE \textbf{Compute} Mining Fee via  RT$_{m}$ 
\STATE M$_{x_{f}}$ = F$_p(j) * ~$RT$_{m}$

\STATE \textbf{Compute} Social Welfare of Seller 
\STATE P$_{f_{j}}$ = F$_p(j) - $~[T$_{x_{f}}$ + M$_{x_{f}}$]
\STATE $SW_{s(i)}$ = P$_{f_{j}} - a_j$

\ENDFOR

\vskip 1.2mm
(3)~\underline{Selecting Miner and Computing Reward}
\vskip 1mm

\STATE \textbf{Collect} M$_{x_{f}}$ Values from Mining Pool \\
\STATE \textbf{Collect} Energy Trading Values for Each VPP\\

\STATE ${P_r}_v$ = [] //Making an empty string

\STATE $\mathscr{S}_{sum}$ = $\sum_{i = 1}^{V_N} {(\mathscr{S}_{i})}$

\FOR {$k \gets$ 1 to $V_N$}
\STATE $Vpp_{PR}(k)$ = $\left(\frac{{\mathscr{S}_{k}}}{\mathscr{S}_{sum}}\right) * 100$
\STATE ${P_r}_v (append)$ = $Vpp_{PR}(k)$
\ENDFOR
\STATE D$_{Vpp} (append)$ = D$_{Vpp}$ \& ${P_r}_{Vpp}$

// \textbf{Select} Winning Miner

\STATE W$_{Vpp} \gets $ random[D$_{Vpp}$] w.r.t Probability Distribution

// \textbf{Select} 2$^{nd}$ Miner for Courtesy Reward

\STATE S$_{Vpp} \gets $ random $\left[D_{Vpp}, \nexists (W_{Vpp})\right] $ w.r.t Probability Distribution

// \textbf{Select} 3$^{rd}$ Miner for Courtesy Reward

\STATE T$_{Vpp} \gets $ random $\left[D_{Vpp}, \nexists (W_{Vpp} \& T_{Vpp})\right] $ w.r.t Probability Distribution

\STATE \textbf{Get} Mining Rewards as Input
\STATE MR $\gets$ Mining Reward

\STATE M$_{sum}$ = $\sum_{i = 0}^{M_xf(m)} \left({M_x}_{f}(i)\right)$
\STATE  RW$_{Vpp}$ = MR + [$\left(\frac{70}{100}\right)$  *~M$_{sum}$]  
\STATE  SW$_{Vpp}$ =  $\left(\frac{20}{100}\right)$  *~M$_{sum}$  
\STATE  TW$_{Vpp}$ =  $\left(\frac{10}{100}\right)$  *~M$_{sum}$ 

\textbf{Mine} the Block in the Network

\RETURN W$_{VPP}$, SW$_{VPP}$, TW$_{VPP}$, RW$_{VPP}$, RS$_{VPP}$, RT$_{VPP}$,  
\RETURN SW${_b}$, SW${_s}$, W${_b}$, W${_s}$

\end{algorithmic}
\end{algorithm}


\subsection{Proof of Energy Market (PoEM)}
Proof of Energy Market mechanism can be \snp{defined as a distributed trading and miner selection protocol for energy blockchain which can be used to carry out energy trading in a decentralized environment, where VPPs act as authoritative nodes. PoEM mechanism can be divided} into three major parts: ($i$) Carrying out Double Auction, ($ii$) Computing transaction fee for each microgrid transaction, and ($iii$) \snp{Selecting winning miner with respect to traded energy and computing mining reward.} \\
In the first step, a double auction is carried out among all the participating nodes, in this step, microgrids and energy buyers have a choice via which they can link themselves with the VPP of their choice, however, they can do it within a specific range specified by the network. \snp{In this step, asks `$a$' are fetched from energy sellers and buyers submit their corresponding bids `$b$' for the slot ($i$). Auction in PoEM works similar to standard double sided auction where the} highest bidder `$W_i$' wins the slot `$S_{id}$' and pays the price `${P_p}_i$' accordingly. A theoretical analysis about allocation and payment rule of the PoEM algorithm is given below in this section. \\
The second step revolves around computation of transaction fee (T$_{x_{fee}}$), mining fee (M$_{x_f}$), and social welfare ($SW_s$) for each microgrid transaction. These values \snp{are calculated according to the prescribed procedures in a way that it benefits all participating parties to a certain extent.} The ratio for transaction fee (RT$_{tx}$) and mining fee (RT$_m$) is decided by mutual agreement between buyers, sellers, and the VPPs. Transaction fee is sent directly to corresponding VPP and mining fee is stored to be sent to the winning miner. \\
In the third step, the mechanism first accumulates all energy values for every individual VPP in order to make a data string for all VPPs. These values are appended in a probability vector ($P_{r_v}$) and a complete database which has information about each VPP and the energy they traded in a specific round (e.g., hourly) is formed. Afterwards, the winner VPP is chosen on the basis of energy it has traded. For example, a VPP has traded 70\% of total energy of the system, then it has 70\% chances of getting selected as a mining VPP in order to get a reward. After selection of winning VPP, second and third winners are chosen for a courtesy reward of 20\% and 10\% accordingly. Afterwards, the block is mined and disseminated to every blockchain node for verification and storage.
\subsubsection{Allocation Rule}\label{allocationsec}
We use the core concepts of double sided auction in our VPP energy trading model~\cite{spref23}. For example, the allocation rule signifies that the highest bidder wins the specific energy slot if and only if, the highest bid is greater than the ask of the seller for that specific slot~\cite{poem01}. A basic formula for single item double auction can be demonstrated as follows:
\begin{equation}
\label{poemeqn01}
\small
X_i(E) = \left[ \underset{b \in b(n)}{argmax} \sum_{j = 1}^n {b_j(E)} \iff \exists \left[a_i(E) \leq b_i(E)\right] \right]
\end{equation}

In the above equation $b_j$ is the bid for $j^{th}$ buyer, and $a_i$ is the ask for $i^{th}$ seller. The equation states that $j^{th}$ bidder can win the bid \textit{if and only if}, his bid is larger than all other bids along with being more than ask of the buyer. Similarly, for multiple energy sellers in a VPP environment, Eqn.~\ref{poemeqn01} can be presented as:

\begin{equation}
\label{poemeqn02}
\footnotesize
\sum_{i = 1}^{S_{max}} X_i(E) = \\
\left[ \sum_{i = 0}^{S_{max}} \left( \underset{b \in b(n)}{argmax} \sum_{j \in n}{b_j(E)} \iff \exists \left[a_i(E) \leq b_i(E)\right] \right) \right]
\end{equation}

\subsubsection{Pricing Rule}

In PPoEM, final payment is decided using differentially private privacy protection mechanism. However, in PoEM, buyer will be paying the amount equal to the bid, so the payment rule is as follows:

\begin{equation}
\label{poemeqn03}
    P_j(E)= 
\begin{cases}
    b_j(E),& \text{if } b_j(E)\geq a_i(E)\\
    0,              & \text{otherwise}
\end{cases}
\end{equation}

In the above equation, $b_j(E)$ is the bid of $j^{th}$ buyer and $a_i$ is the ask for $i^{th}$ seller in a condition that ask is always greater than the bid.

\subsubsection{Miner Choosing Phenomenon}
\snp{In our PoEM algorithm, a miner is chosen on the basis of energy it has traded in the previous round.} A miner is chosen with respect to the ratio of energy it has traded. First of all, the data of all traded energy in a specific round is collected and a parameter of energy sum `$\mathscr{S}_{sum} = \sum_{j = 1}^{V_N} \sum_{i=1}^n \left(E_{V_{i}}(i))\right)$' is calculated by accumulating energy values from all VPP miners. Afterwards, an intermediary vector `$\vec{P_{rv}}$' is used to calculate final distribution `D$_{vpp}$' as follows:


\begin{equation}
\label{poemeqn06}
\vec{P_{rv}} = \vec{ P_{rv}} \frown \sum_{k = 1}^{V_N} \left[\frac{\mathscr{S}_k}{\mathscr{S}_{sum}}*100\right]
\end{equation}
\begin{equation}
\label{poemeqn10}
D_{vpp}  = D_{vpp} \frown \vec{P_{rv}}
\end{equation}
From the above distribution, first, second, and third winning VPP is computed via random selection phenomenon. First VPP will mine the block and get the major reward, however, the second and third VPP gets courtesy reward accordingly. 
\begin{equation}
\label{poemeqn07}
\small
\begin{cases}
    W_{vpp}  = Rand[D_{vpp}],& $Winning VPP$\\
    S_{vpp}  = Rand[D_{vpp}, \nexists~W_{vpp}],& $Second winner$\\
T_{vpp}  = Rand[D_{vpp}, \nexists~W_{vpp}~\&~S_{vpp}],& $Third winner$\\

\end{cases}
\end{equation}

\subsubsection{Mining Reward Calculation}
\snp{Mining reward in our PoEM mechanism mainly depend upon two factors, one is fixed mining reward ($MR$) which is given by the governing authority and has a fixed value, and second factor is mining fee ($M{_{x_f}}$). This mining fee is deducted at every energy trading transaction of microgrids carried out via VPPs. The amount is accumulated as mining sum ($M_{sum}$) at the mining pool in the form of VPP coin and is distributed at the end of the mining process. The formula for calculation of mining reward is as follows:}
\begin{equation}
\label{poemeqn08}
\small
\begin{cases}
    RW_{vpp}  = MR + (0.7*M_{sum}),& $Winning Reward$\\
    RS_{vpp}  = (0.2*M_{sum}),& 2^{nd}~$Courtesy Reward$\\
RT_{vpp}  =(0.1*M_{sum}),& 3^{rd}~$Courtesy Reward$\\

\end{cases}
\end{equation}
These ratios can be varied and can be decided after discussion between VPPs and the controlling nodes. However, just for the sake of simplicity, we fixed these ratios in our algorithm.

\subsubsection{Social Welfare}
In a sealed bid double auction, social welfare can be termed as the utility of participants with respect to their bids and asks~\cite{poem02,spref21}. In the PoEM algorithm, only the social welfare of sellers is computed because buyers will be paying the amount they bid for a specific slot. \snp{The formula to calculate social welfare of $i^{th}$ seller in presence of $j^{th}$ buyer is as follows:} 
 \begin{equation}
\label{poemeqn09}
SW_{s(i)}  = P_{f_i} – a_j
\end{equation}


\begin{algorithm}[htp]\scriptsize
\caption{Private Proof of Energy Market (PPoEM) Algorithm}
\label{algoDP2}
\textbf{Input:}  b, a, $\varepsilon_1$, $\varepsilon_2$, $\varepsilon_3$ E, S, $\mu$, S$_v$, S$_C$

\textbf{Output:} W$_{VPP}$, SW${_b}$, SW${_s}$, W${_b}$, W${_s}$

\vskip 2mm
\hspace*{4mm} (1) \underline{Carrying out Private Double Auction}
\vskip 1mm

\begin{algorithmic}[1]

\STATE \textbf{max$_{bid_{(s)}} \gets argmax[sort(b)]$}
\FOR {\textbf{each}~\textit{seller} j $\gets$ 1 to $S_{max}$}

\FOR {\textbf{each}~\textit{buyer} k $\gets$ 1 to $N_{max}$}

\IF {(b $\geq$ a \& b == max$_{bid}(j)$)}
\STATE \textbf{Calculate} $j^{th}$ energy slot winner $({W}_j)$ w.r.t rule of allocation
\STATE $W_j(x) = argmax_b \sum_{k \in N}{X_k(b)}$\\
//~$W_j(x)$ is the selected winner for slot $E(j)$
\ELSE 
\STATE \textbf{return} 'bid did not match the ask'
\STATE \textbf{break;}
\ENDIF

\textbf{Calculating} Differentially Private Price From Here
\STATE W$_{bid}(j) \gets$ Winning bid from $j^{th}$ buyer
\STATE $a(j) \gets$ Seller Ask for that Specific Slot
\STATE dif = W$_{bid}(j)$ - $a(j)$
\STATE Laplace Mean = $\frac{dif}{2}$

\STATE \textbf{Compute} DP Price String via Lap($W_{bid}(j)$, F$_i$, $\varepsilon_1$) Store in P$_v$

\textbf{Select} Differentially Private Price via Exponential Mechanism
\STATE P$_w \gets$ Winner Probability Distribution
\STATE $\Delta$q$ \gets$ S$_v$
\STATE $P_w(F(P_v, q_1, O_p) = o_p) \gets$ \newline
\vskip 0.3mm
\hspace*{8mm} \normalsize $\frac{\exp (\frac{\varepsilon_2 . q_1(P_v, o_p)}{2 \Delta q_1})}{\sum_{{o_p}^\prime \in O_p} \exp (\frac{\varepsilon_2 . q_1(P_v, {o_p}^\prime)}{2 \Delta q_1})}$\\
\scriptsize
\textbf{Pick} Final Random Price (F$_p(j)$) from P$_w$ Probability Distribution 
\STATE F$_p(j) \gets$ random(P$_w$)

\ENDFOR

\STATE \textbf{Append} Winner ID, price, asks for that slot, energy slot
\STATE \textbf{Append} W$_b$[I$_d$, F$_p(j)$, $a(j)$, E$_s$]
\ENDFOR

\vskip 1.2mm
(2)~\underline{Compute Social Welfare, Transaction Fee \& Mining Fee}
\vskip 1mm
\STATE RT$_{tx}$, RT$_{m} \gets$ via $S_s$ from Algorithm 1
\FOR {$j \gets$ 1 to $W_{s(max)}$}
\STATE \textbf{Compute} Transaction Fee via  RT$_{tx}$ 
\STATE T$_{x_{f}}$ = F$_p(j) * ~$RT$_{tx}$

\STATE \textbf{Compute} Mining Fee via  RT$_{m}$ 
\STATE M$_{x_{f}}$ = F$_p(j) * ~$RT$_{m}$

\STATE \textbf{Compute} Social Welfare of Seller 
\STATE $SW_{s(i)}$ = F$_p(j) - $~[T$_{x_{f}}$ + M$_{x_{f}}$] $-~a_j$
\STATE $SW_{b(i)}$ = $b(i) - $ F$_p(j)$

\ENDFOR

\vskip 1.2mm
(3)~\underline{Selecting Miner and Computing Reward}
\vskip 1mm
\textbf{Get} List \& Energies of all Participating VPPs\\
\STATE \textbf{Generate} Probability Distribution of all Energies w.r.t $\varepsilon_2$ differential privacy
\FOR {j $\gets$ 1 to $V_N$}
\STATE VPP$_{pr} \gets $ Prob distribution of V$_{pp}$ Energy
\STATE $\Delta q_2 \gets$ M$_s$
\STATE VPP$_{pr}(F(L_v, q_2, D_{vpp}) = d_{vpp}) \gets$ \newline
\vskip 0.3mm
\hspace*{8mm} \normalsize $\frac{\exp (\frac{\varepsilon_3 . q_2(L_v, d_{vpp})}{2 \Delta q_2})}{\sum_{{d_{vpp}}^\prime \in D_{vpp}} \exp (\frac{\varepsilon_3 . q_2(L_v, {d_{vpp}}^\prime)}{2 \Delta q_2})}$\\

\scriptsize \STATE P$_{RV}$(\textbf{Append}) = VPP$_{pr} (i)$
\ENDFOR

\STATE $D_{vpp}$(\textbf{Append}) = $D_{vpp}$ \& P$_{RV}$

\textbf{Select} Mining Node w.r.t Probability Distribution in $D_{vpp}$

// \textbf{Select} Winning Miner

\STATE W$_{Vpp} \gets $ random[$D_{vpp}$] w.r.t Differential Privacy Distribution

// \textbf{Select} 2$^{nd}$ Miner for Courtesy Reward

\STATE S$_{Vpp} \gets $ random $\left[D_{vpp}, \nexists (W_{Vpp})\right] $ w.r.t Differential Privacy Distribution

// \textbf{Select} 3$^{rd}$ Miner for Courtesy Reward

\STATE T$_{Vpp} \gets $ random $\left[D_{vpp}, \nexists (W_{Vpp} \& T_{Vpp})\right] $ w.r.t Differential Privacy Distribution

\STATE \textbf{Get} Mining Rewards as Input
\STATE MR $\gets$ Mining Reward

\STATE M$_{sum}$ = $\sum_{i = 0}^{M_xf(m)} \left({M_x}_{f}(i)\right)$

\STATE \textbf{Pick} Random Number

\STATE R$_R \gets$ random$\left(0~to~M{_{sum}}\right)$ // This is Winner Reward
\STATE R$_o $ = M$_{sum}$ - R$_R$
 
\STATE  RW$_{Vpp}$ = MR + R$_R$  
\STATE  SW$_{Vpp}$ =  $\left(\frac{70}{100}\right)$  * R$_o $  
\STATE  TW$_{Vpp}$ =  $\left(\frac{30}{100}\right)$  * R$_o $ 

\textbf{Mine} the Block in the Network

\RETURN W$_{VPP}$, SW$_{VPP}$, TW$_{VPP}$, RW$_{VPP}$, RS$_{VPP}$, RT$_{VPP}$,  
\RETURN SW${_b}$, SW${_s}$, W${_b}$, W${_s}$, T${_K}_{sum}$

\end{algorithmic}
\end{algorithm}


\subsection{Private Proof of Energy Market (PPoEM)}

In order to develop the PPoEM mechanism from PoEM, we integrate differential privacy at two places in the PoEM mechanism. Firstly, both Laplace and Exponential mechanism of differential privacy are used to carry out differentially private price selection in the double auction process, which preserves bid privacy in a sealed bid auction. Afterwards, the Exponential mechanism of differential privacy is used to carry out differentially private mining selection, which is the core part of our private mining algorithm. In the following subsection, we discuss the above mentioned differences in PPoEM algorithm from a technical perspective. \\
\subsubsection{Differentially Private Pricing Rule}
The allocation rule of the PPoEM algorithm is the same as that of PoEM algorithm, therefore, we only discuss the pricing rule here. In PPoEM algorithm, the final price is calculated from the winning bid in a differentially private manner by keeping in view the social welfare of both the buyer and seller. First of all, the difference between the winning bid and ask is calculated ($dif = W_{bid(j)} – a(j)$) in order to determine the price string limitations. Afterwards, Laplace differential privacy mechanism is used to determine the price string in between the ask and the final price. The length of string ($l$) can be adjusted according to the requirement and privacy conditions. The random pricing values are calculated and appended in a vector called $\vec{P_v}$ as follows:
\begin{equation}
\label{ppoemeqn01}
\vec{P_v} = \vec{P_v} \frown \sum_{i=1}^{l} Lap\left(W_{bid}(j), F_i, \varepsilon_1\right)
\end{equation}

Afterwards, a differently private price is selected using Exponential mechanism as follows:


\begin{dmath}
\label{ppoemeqn02}
\small
P_w(F(P_v, q_1, O_p) = o_p) \gets \frac{\exp (\frac{\varepsilon_2 . q_1(P_v, o_p)}{2 \Delta q_1})}{\sum\limits_{{o_p}^\prime \in O_p} \exp (\frac{\varepsilon_2 . q_1(P_v, {o_p}^\prime)}{2 \Delta q_1})}
\end{dmath}

In the above equation, $\Delta q_1$ is the sensitivity value, which can be varied according to the requirement. After successfully calculation and appending pricing values in the distribution. A random value is picked from $P_w$, which serves the purpose of the final price. It is ensured that the price is always greater than the ask and less than the bidding value.

\subsubsection{Differentially Private Miner Selection}
Miner selection in our PPoEM mechanism is carried out using the Exponential mechanism of differential privacy~\cite{spref27}. Different from PoEM, instead of calculating energy ratios, we calculate energy probabilities in an exponentially private manner, and then we choose the winning miner from that probability distribution. First of all, energy values of all VPPs are fed as an input to the Exponential mechanism, which calculates the probability of selection for each VPP according to the chosen sensitivity and privacy parameter. Higher the value of $\varepsilon_3$, higher is the chances of selection of VPP with maximum energy trading. The formula for differentially private miner section is given as follows:

\vspace{-0.4em}
\begin{dmath}
\label{ppoemeqn03}
\footnotesize
$VPP$_{pr}(F(M) = d_{vpp}) \gets \frac{\exp (\frac{\varepsilon_3 . q_2(L_v, d_{vpp})}{2 \Delta q_2})}{\sum\limits_{{{d_{vpp}}^\prime \in D_{vpp}}} \exp (\frac{\varepsilon_3 . q_2(L_v, {d_{vpp}}^\prime)}{2 \Delta q_2})}
\end{dmath}

In the above equation, $M$ = $(L_v, q_2, D_{vpp}) $ and $q_1$ is the sensitivity value, which can be varied according to the requirement. We carry out experiments at different $\varepsilon_3$ values in order to demonstrate the functioning from a technical perspective. A detailed discussion about implementation and evaluation is provided in the Section~\ref{perfsection}. After the successful calculation of mining distribution, the first, second, and third miner is chosen similar to the PoEM mechanism as mentioned in Eqn.~\ref{poemeqn07}.

\subsubsection{Private Miner Reward}
In order to make the miner reward more confidential, we picked a random reward value between $0~to~M_{sum}$ and named it as $R_R$. After calculation of ($R_R$), a parameter called the remaining reward ($R_o$) is calculated by subtracting the value from mining sum ($R_o$ = $M_{sum}  - R_R$). This value of $R_o$ is used to calculate the courtesy reward for second and third VPP as follows:
\begin{equation}
\label{ppoemeqn008}
\small
\begin{cases}
    RW_{vpp}  = MR + R_R,& $Winning Reward$\\
    RS_{vpp}  = (0.2*R_o),& 2^{nd}~$Courtesy Reward$\\
RT_{vpp}  =(0.1* R_o),& 3^{rd}~$Courtesy Reward$\\

\end{cases}
\end{equation}

\subsubsection{Social Welfare Maximization}

\snp{In PPoEM mechanism, the social welfare is maximized for both participants in order to motivate them to participate in the auction. The formulas for calculation of social welfare of $i^{th}$ seller and $j^{th}$ buyer are as follows:}
\begin{equation}
\label{ppoemeqn07}
SW_{s(i)}  = P_{f_i} – a_j 
\end{equation}
\begin{equation}
\label{ppoemeqn08}
SW_{b(j)}  = b(i) - P_{f_i} 
\end{equation}

\subsection{Functioning, Operation, \& Integration Details in VPP} \label{function}

This section discusses the functioning of VPT energy trading blockchain network in detail from the point of view of block generation, validation, and VPP coin.
 \subsubsection{Block Generation}
\snp{In our VPT model, block generation is carried out right after choosing the winning miner. The selected leader/winning miner performs this step} in order to win the mining reward. Furthermore, in the leading time-period, the leader/winning VPP can pick transactions from invalidated portions of the mining pool and can validate the transactions to get extra reward. For example, from each transaction validation, he gets some percentage from the transaction fee of trading VPP. 
\subsubsection{Block Validation}
Firstly, all transactions and the complete block contents are validated by the leader VPP, and afterwards it is disseminated to all VPPs for further validation. Afterwards, all VPPs act as validators and validate the block in order to confirm its integrity. VPPs generate block hash via SHA-256 algorithm and compare the newly generated hash with the received hash. If both hash values match, then the block is considered as a legal block and is then forwarded for the updating of ledger. Microgrids and other participating blockchain nodes only act as a viewer and cannot validate the block, rather they can just view the contents of the block after successful dissemination and approval.

\subsubsection{VPP Coin}
In order to carry out efficient and timely trading, and in order to reduce intermediary banks from our network, we introduce the concept of VPP coin. The aim of our VPT mechanism is to enhance energy trading rather than carrying out crypto-trading, therefore, participating nodes (such as microgrids) cannot trade/exchange VPP coins with each other. There are only three use cases for participants; firstly, they can only earn coins by selling their energy, secondly, they can only  spend the VPP coin by purchasing energy. Finally, if they want to purchase or sell VPP coins in return of local currency, they can only do it via authoritative nodes. 

\begin{figure*}[t]
\centering
\captionsetup{justification=centering}
\begin{center}

\subfigure[]{
\includegraphics[scale = 0.25]{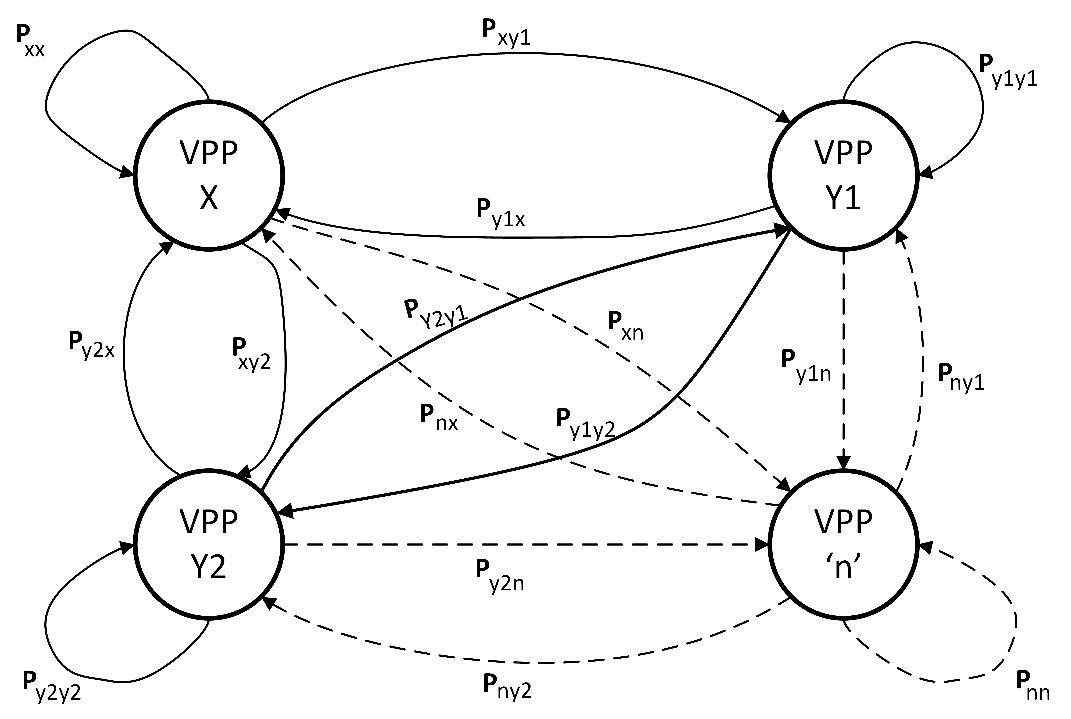}
}
\subfigure[]{
\includegraphics[scale = 0.25]{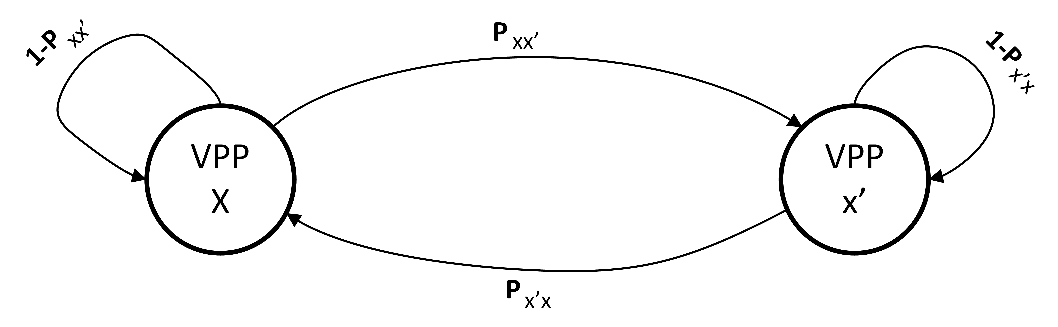}
}
\subfigure[]{
\includegraphics[scale = 0.25]{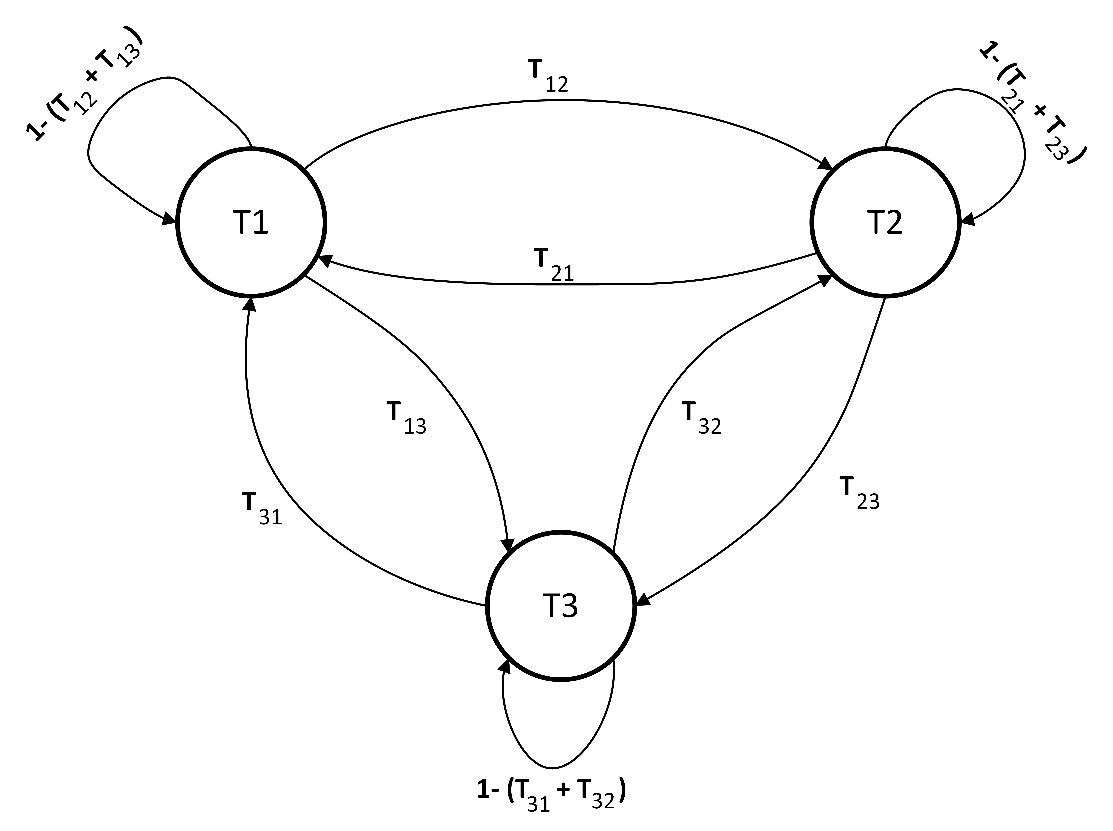}
}

\caption{\textsc{Markovian State Probabilities for VPT}   (a) Market Capture Probability Containing '$n$' VPPs \newline (b) Market Capture Containing two VPPs ($x$ and $x'$) (c) State transition diagram for VPPs till next mining election }
\label{fig:MarkovGraph}
\end{center}
\end{figure*}





\section{\snp{Security, Privacy, and Functionality Analysis}}
In this section, we carry out \snp{analysis of our VPT model for various functionalities such as privacy, security, VPP market capture, market race,} market expectations, etc., along with discussing complexity analysis and other theoretical aspects. 

\subsection{Differential Privacy Analysis}\label{DPAnalysis}
Our proposed PPoEM algorithm uses the concept of differential privacy to protect buyers bidding values and energy trading values of VPP. In order to prove that PPoEM algorithm follows differential privacy guarantees, we provide extensive theoretical analysis of it given in the following discussion.

\begin{Lemma}{Consider $X_1(q)$ and $X_2(q)$ be two differentially private algorithms with privacy budgets $\varepsilon_1$ and $\varepsilon_2$ respectively. Then, $X(q) = (X_1(q),X_12(q))$ satisfies ($\varepsilon_1 + \varepsilon_2$)-differential privacy according to composition theorem~\cite{dpbook}. }
\label{lemmalabel01}
\end{Lemma}

\vskip 1mm
\begin{theorem}{\textit{Laplace Mechanism in Price Selection of PPoEM Algorithm is $\varepsilon_1$-differentially private.}}\\
\label{difPriv01}
\hspace{10mm}\textit{\textbf{Proof:}} See Appendix for Proof
\end{theorem}

\begin{theorem}{\textit{Exponential price selection and miner selection phenomenon of our PPoEM mechanism provides $\varepsilon_2$-differential privacy and $\varepsilon_3$-differential privacy respectively.}}\\
\label{difPriv02}
\hspace{10mm}\textit{\textbf{Proof:}} See Appendix for Proof
\end{theorem}

\begin{theorem}{\textit{Differentially private auction of PPoEM satisfies $\varepsilon$-differential privacy.}}\\
\label{DPComplete}
\hspace{10mm}\textit{\textbf{Proof:}} See Appendix for Proof
\end{theorem}
\vspace{-1mm}
\subsection{\snp{Security Analysis}}\label{SecAnalysis}
\snp{Our proposed VPT model has an ability to carry out defence against various traditional security attacks due to usage of basic primitives of cryptography in the blockchain (e.g., symmetric and asymmetric encryption via keys). Similarly, an adversary will not be able to carry out various attacks such as inference, replication, forgery, etc due to added digital signature and differential privacy in it. In this analysis, we carry out analysis from the perspective of certain security requirements in blockchain based energy trading systems.}

\begin{theorem}{\textit{\snp{Our proposed VPT model ensures wallet security, transaction authenticity, block confidentiality, block integrity, blockchain data availability, over provisioning resilience, and efficient fork resolution.}}}\\
\label{securityanalysis}
\hspace{10mm}\textit{\textbf{Proof:}} See Appendix for Proof
\end{theorem}

\begin{theorem}{\textit{\snp{Our proposed VPT model provides effective resillience to sybil and inference attacks.}}}\\
\label{attackanalysis}
\hspace{10mm}\textit{\textbf{Proof:}} See Appendix for Proof
\end{theorem}

\vspace{-1mm}
\subsection{Market Capture Probability}
We divide market capturing into two different probabilities named as market race and steady market probability, which are given in the further sections.
\subsubsection{Market Race Probability}
Consider a VPP network with $V_N$ number of VPPs participating in energy trading process with different amount of traded energy till reported time $i$. \snp{Consider a VPP $x$, which traded maximum amount of energy till the end of election time-out time (e.g., one hour for hourly mining). The probability that this VPP $x$ was always ahead of the second highest VPP $y$ is given in the following theorem.}

\vskip 1mm
\begin{theorem}{\textit{The probability that winning VPP ($x$) was always ahead of second highest VPP is given by}}
\label{ProbT01}

\begin{equation}
\label{Teqn01}
\mathcal{P}_{\mathscr{S}_{x}, \mathscr{S}_{y}} = \frac{\mathscr{S}_{x}(\mathscr{S}_{x}-E_v(i)) - \mathscr{S}_{y}(\mathscr{S}_{y} + E_v(i))    }{\mathscr{S}_{x}(\mathscr{S}_{x} - E_v(i)) + \mathscr{S}_{y}(\mathscr{S}_{y} - E_v(i)) + 2\mathscr{S}_{x}\mathscr{S}_{y}}
\end{equation}

\hspace{10mm}\textit{\textbf{Proof:}} See Appendix for Proof

\end{theorem}

\subsubsection{Steady Market Probability}

Similarly, when VPPs \snp{attract selling prosumers by providing them} incentives based upon their energy, timing, power factor, etc. Then transition of customers occur between VPPs in a way that some microgrids from one VPP $x$ move to other VPPs to get better incentives, and similarly, \snp{some prosumers from other VPPs move to $x$ VPP for better incentives.} This complete system forms an $n$ state aperiodic Markov chain, similar to the one that can be analysed in Fig.~\ref{fig:MarkovGraph}(a). The model given in the figure can further be reduced to form a two state Markov model for a VPP $x$ and other VPPs $x'$, which describe the transition of customers among one VPP and all other VPPs (as given in Fig.~\ref{fig:MarkovGraph}(b)). This further leads to a Markovian problem of VPP $x$ capturing the certain proportion of the market at a certain time interval in presence of some specific transition probabilities, which is evaluated in the Theorem~\ref{ProbT02}.
\vskip 1mm
\begin{theorem}{\textit{The steady state market capture probability for a VPP $x$ is given by  }}\newline
\label{ProbT02}
\begin{equation}
\label{Teqn05}
C_x = \frac{\mathcal{P}_{x'x}}{\mathcal{P}_{x’x}+\mathcal{P}_{xx’}}
\end{equation}

\hspace{10mm}\textit{\textbf{Proof:}} See Appendix for proof












\end{theorem}


\begin{table}[H]
\begin{center}
 \centering
 \footnotesize
 \captionsetup{labelsep=space}
 \captionsetup{justification=centering}
 \caption{\textsc{\\State Transition Probabilities for VPPs Till Next Mining Election.}}
  \label{tab:STP01}

\begin{tabular}{|c| c c c|}

\hline
\textbf{State} & ~ &  $\textbf{P}_{\textbf{E}_k}\textbf{(x)}$ ~ &~\\ \cline{2-4}
\textbf{Space} & \textbf{0-10 \% } & \textbf{ 10-20\% } & \textbf{ 20-100\% }\\ \hline

\textbf{T1} &  T3 & T2 & T1 \\ \hline

\textbf{T2} & T3 & T2 & T1 \\ \hline

\textbf{T3} & T3 & T2 & T1 \\ \hline

 \end{tabular}
  \end{center}
\end{table}


\subsection{Winning State Probability}

In order to model the behaviour of winning miners, we divide the total energy traded into three states $T = {T_1,~T_2,~\&~T_3}$, with $T_1$ being the miners having a high probability of winning the next election. The transition of VPPs among these states can be modelled as an irreducible aperiodic Markovian chain because of dynamic transitions, given in Fig.~\ref{fig:MarkovGraph}(c). Transition among these VPPs is carried out according to the rules given in Table~\ref{tab:STP01}.

\subsubsection{VPP Winning State Probability}

For a VPP, it is important to be in the highest winning probability state $T_1$ for most of the time during the trading period in order to maximize its chance of winning. The chances of a VPP winning a minor election while being in state $T_2$ \& $T_3$ is fairly less as compared to one being in $T_1$. Therefore, we consider these states as low winning states. From now onwards, we will derive the rate of transition and stay from high winning state $T_1$ as compared to that of low probability states $T_2$ \& $T_3$.
\vskip 1mm
\begin{theorem}{\textit{The average time length in which a VPP remains in high probability winning state $T_1$ is:}}
\label{ProbT05}

\begin{equation}
\vec{WT} = \frac{\sum_{j\in T_1} (\pi_j)}{\sum_{l\in T_3}\sum_{k\in T_2}\sum_{j \in T_1} \pi_j (P_{jk} + P_{jl})}
\end{equation}

\hspace{10mm}\textit{\textbf{Proof:}} See Appendix

\end{theorem}



\subsection{Prospective Profit During Leading Time}
When a VPP wins an election, he becomes an elected leader till the next election. During this time period, it can pick the invalidated transactions from the mining pool and validate them for the next block. In this way, it can earn extra profit during its leading time. In order to monitor the prospective profit that a \snp{VPP can make during its reign, we model it with a queuing approach discussed in the next theorem.}

\vskip 1mm
\begin{theorem}{\textit{The prospective profit that a VPP can make during his leading period is:}}
\label{LeadTheorem}
\vspace{-0.5em}
\begin{equation}
\label{LeadEqn01}
Total Profit = T_p = \frac{R_A M \left[1 - \left(\frac{R_A}{R_s}\right)^{T_L}\right]}{1-\left(\frac{R_A}{R_s}\right)^{T_L+1}} ~–~ C R_s
\end{equation}

\hspace{10mm}\textit{\textbf{Proof:}} See Appendix

\end{theorem}

\subsection{Complexity Analysis}
Our proposed VPT energy trading and mining model is computationally efficient from the perspective of time and power. This is because of the reasons that a lower possible number of iterations are considered to carry out double auction and miner selection processes.

\subsubsection{Computational Complexity}
From the perspective of computational complexity, PoEM comprises three major parts that can be executed independently by providing required inputs depending upon the requirement. It is also important to note that the complexity of PPoEM algorithm (Algorithm~\ref{algoDP2}) does not have significant differences as compared to complexity analysis of PoEM except for integration of exponential differential privacy steps. Therefore, to provide a broader picture of our proposed mining mechanism, we only calculate the complexity of the PoEM algorithm, which can easily be linked with PPoEM in case of need.

\vskip 1mm
\begin{theorem}{\textit{The computational complexity of auction part in our PoEM algorithm is upper bounded by \newline $\mathcal{O} (max{N\log(N), SN})$.}}\newline
\label{theorem01}

\hspace{10mm}\textit{\textbf{Proof:}} See Appendix for proof.

\end{theorem}

\vskip 1mm
\begin{theorem}{\textit{The upper bound computational complexity of Social welfare computation, transaction fee calculation, and mining fee determination is $\mathcal{O}(S)$.}}\newline
\label{theorem02}
\hspace{10mm}\textit{\textbf{Proof:}} See Appendix for proof.

\end{theorem}

\vskip 2mm
\begin{theorem}{\textit{Miner selection and reward computation part of PoEM has an upper bound computational complexity of $\mathcal{O}(Ws_{max})$.}}\newline
\label{theorem03}
\hspace{10mm}\textit{\textbf{Proof:}} See Appendix for proof.

\end{theorem}

Keeping in view the complete analysis, the computational complexity of all three parts of PoEM algorithm can be summarised as $\mathcal{O}(max({N\log(N), SN})$ + $\mathcal{O}(S)$ + $\mathcal{O}(Ws_{max}))$. In which the most dominant part is carrying out double auction having the worst computational complexity of $\mathcal{O}(SN) \approx \mathcal{O}(n^2)$ in case when $S \approx N$ and $\mathcal{O}(SN) \gg \mathcal{O}(N\log(N))$ after break-even point.


\subsubsection{Power Consumption}

The proposed VPT model can be deployed at any VPP without \snp{having the trouble about power consumption.} Firstly, the trading will take place at least after one hour, therefore, the possibility of bottleneck is near to minimum. Secondly, the \snp{proposed VPT model has low memory} and computational complexity as compared to other traditional consensus variants that use mining difficulty for choosing a miner. Finally, VPPs are also equipped with strong infrastructure to carry out various tasks such as blockchain management~\cite{relnew01}, infrastructure load management~\cite{spref14}, and communication via IEC61850~\cite{analysis02}.


\begin{algorithm}[H]\scriptsize
\caption{Recovery Response for VPT}
\label{algoRecovery}

\textbf{Input:}  $V_N$, $W_{VPP}$, $X$

\textbf{Output:} Recovery request resolved

\begin{algorithmic}[1]

\STATE \textbf{Begin:\newline}
//\textit{Recovery} message launched by '$X$' node

\IF {($X$ node is VPP)}
\STATE $M_{RN} = \frac{2}{3} V_N$ \newline
//\textit{X} receives chain \& hash from active VPPs
\STATE $H_M$ = Majority\{Hash$_{V_{1}}$, Hash$_{V_{2}}$, ....., Hash$_{V_{N}}$\}
\STATE $C_{H_M}$ = Count\{Occurrence of $H_M$\}

\IF {($C_{H_M} \geq M_{RN}$)}
\STATE \textbf{Update} current chain state w.r.t $H_M$
\ELSE
\STATE \textbf{Send} Warning Message to Leader VPP ($W_{VPP}$)
\ENDIF

\ELSIF {($X$ node is VPP)}
\STATE \textbf{Request} directly sent to Leader VPP ($W_{VPP}$)
\STATE \textbf{X Receives} the chain from $W_{VPP}$, and update its status
\ENDIF

\RETURN Recovery Request Resolved

\end{algorithmic}
\end{algorithm}


\subsection{Recovery Response}
In case if a node fails and recovers from the crash, it is compulsory for the node to go through the complete recovery process before continuing the normal operations. The first step in the recovery response is launching the ‘\textit{Recovery}’ message, which transmits the message to all other VPP nodes in case if a VPP is recovering, and only send the message to leader VPP ($W_{VPP}$) node in case of a smart metering/energy trading node being recovered. In case of VPP recovery, the required numbers are calculated using the number of total active VPPs via ($M_{RN} = \frac{2}{3} * V_{N}$). Afterwards, the received hashes are compared and the majority number of hashes are picked from received hashes, and if the majority hash count is greater than the calculated $M_{RN}$, then the state of recovering VPP is updated with respect to the majority hash chain. Elsewise, if $C_{H_{M}} \leq M_{RN}$, then a warning message is launched to leader VPP to troubleshoot the network in order to overcome any inconsistency in the network. Furthermore, if the request launching node is a smart meter or energy trading node, then the request is directly sent to leader VPP $(W_{VPP})$, which sends the current version of the chain to the node to update its status. The detailed pseudo-code for recovery response has been provided in Algorithm~\ref{algoRecovery}.







\begin{figure}[h]        
\centering
\includegraphics[scale = 0.29]{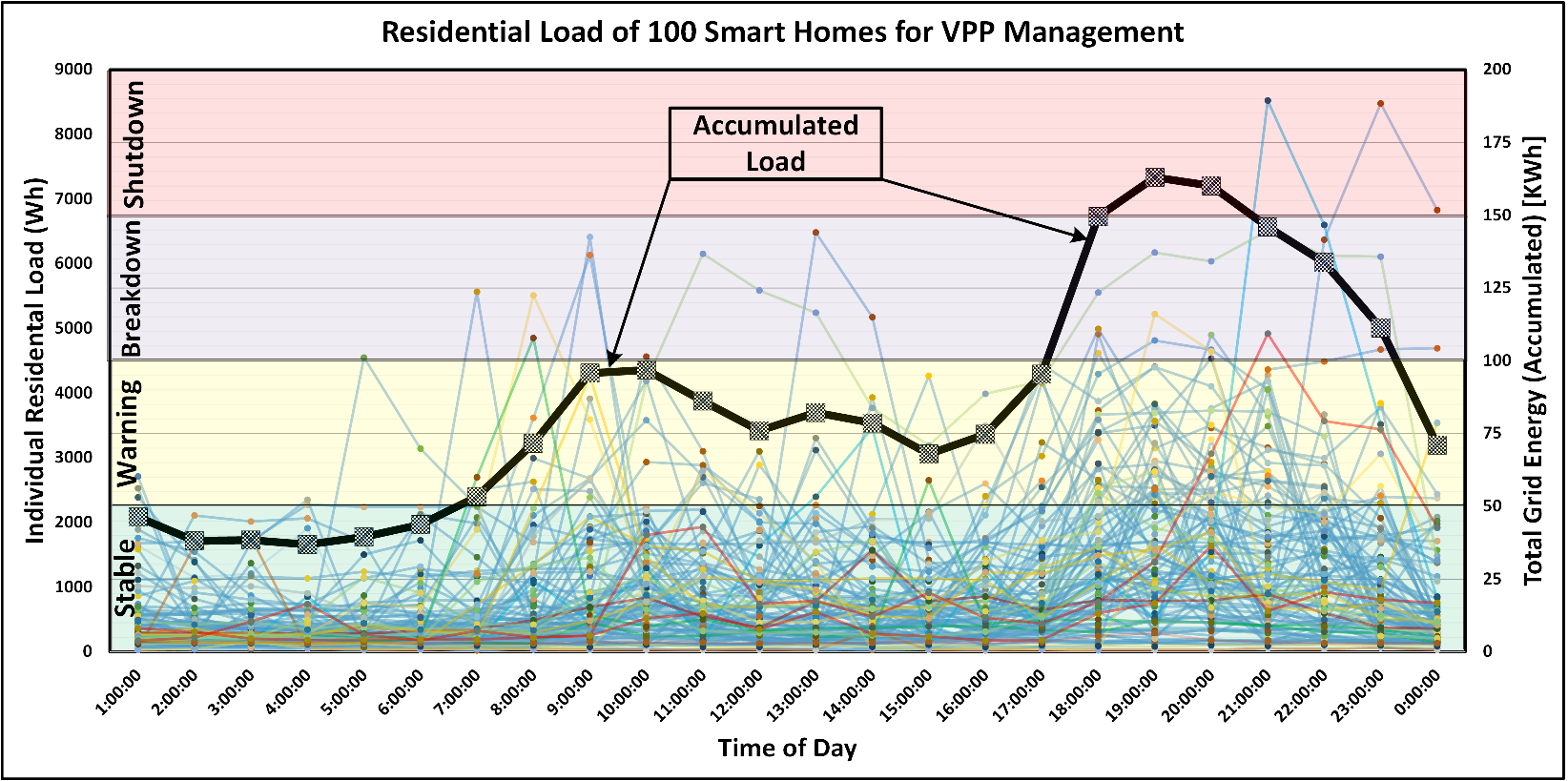}
 \caption{Accumulated Residential Energy Usage of 100 Smart Homes at Different Management Levels}      
  \label{fig:figLoadCurve}   
\end{figure}


\begin{figure}[t]        
\centering
\includegraphics[scale = 0.65]{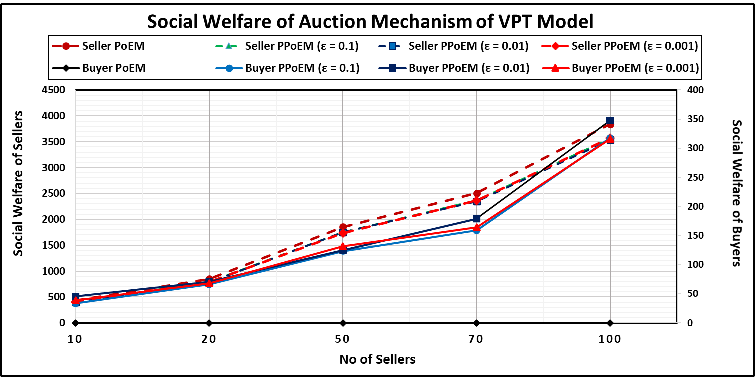}
 \caption{Social Welfare of Auction Mechanisms of VPT Model at Various Parameters}      
  \label{fig:figVPTauction}   
\end{figure}

\section{Performance Evaluation of Virtual Private Energy Trading}\label{perfsection}
\snp{To implement VPT model, we develop the functionalities of traditional and differentially private double sided auction at each VPP via Python. Moreover, to determine system state, we use real-time data of 100 smart homes from the AusGrid dataset of residential profiles~\cite{ausgrid}.} We further develop a decentralized blockchain based model to evaluate PoEM and PPoEM mining functionalities. After successful mining election, a block is formed and this block containing all information regarding the future leader, etc, is then mined to blockchain and is also sent to other validating nodes for validation. 
\subsection{PoEM \& PPoEM Double Auction}
In order to evaluate our VPT trading model, we first evaluate Algorithm~\ref{AlgoSysState} and determine system state on the basis of residential load profiles given in~\cite{ausgrid}. The decision of system state is taken on the basis of accumulated load usage by smart homes. \snp{Afterwards, the mining fee and transaction fee percentage is determined on the basis of system state.} A graphical evaluation of system state determination has been provided in Fig.~\ref{fig:figLoadCurve}. After determining the system state, the transaction fee and mining fee is determined for PoEM and PPoEM elections. This step is carried out to encourage microgrids to sell their stored energy at the time of energy shortfall. After that, PoEM and PPoEM auctions are carried out and social welfare is evaluated for each participating buyer and seller. The evaluation of social welfare on the basis of system state has been provided in Fig.~\ref{fig:figVPTauction}. The given figure shows the trend of increase in social welfare of sellers and buyers with respect to increase in buyers. For example, the social welfare is minimum for PoEM and PPoEM when minimum buyers are participating, and it increases with the increase in number of buyers. \\
For instance, in PoEM, when the number of buyers are 10, the social \snp{welfare of sellers gets around 500, and this} value increases with increase in the number of sellers. Similarly, the social welfare of buyers also increases with the increase in the number of sellers as shown in the given figure. It is important to note that the social welfare of buyers is only applicable to PPoEM algorithm, as in case of PoEM algorithm the final price is the ask of seller, therefore, buyers social welfare is not taken into account in PoEM auction. Furthermore, the auction value at three privacy budgets of ($\varepsilon$ = 0.1, 0.01, \& 0.001) is evaluated for different sellers and it can be seen from the output graphs that PPoEM provides a similar social welfare for sellers along with providing differentially private protection for buyers bids and sellers asks.

\begin{table}[H]
\begin{center}
 \centering
 \scriptsize
 \captionsetup{labelsep=space}
 \captionsetup{justification=centering}
 \caption{\textsc{Comparative Analysis of PoEM and PPoEM with PoA and PoE.}}
  \label{tab:compare}

\begin{tabular}{|p{5em}|p{4.5em}|p{8.5em}|p{4em}|p{4em}|}

\hline

\textbf{Consensus/\newline Mining} &  \centering \textbf{Incentivzing High Trader} & \textbf{Incentive Type} & \textbf{Ledger Storage Privacy} & \textbf{Complexity} \\ \hline

\textbf{PoA~\cite{compare01, compare03, compare04}} &  No & Mining Reward & No & $O(n)$ \\ \hline

\textbf{PoE~\cite{relnew01}} & Partially & Mining Reward & No & $O(n)$ \\ \hline

\textbf{PoEM (proposed)} & Complete & Mining Reward + Tx Fee + Mining Fee & No & $O(n)$ \\ \hline

\textbf{PPoEM (proposed)} & Complete & Mining Reward + Tx Fee + Mining Fee & Yes & $O(n)$ \\ \hline

 \end{tabular}
  \end{center}
\end{table}


\subsection{PoEM \& PPoEM Mining Election}
We divide the VPT mining election evaluation section into two parts, firstly, we discuss mining winner selection and then we discuss incentive determination for PoEM and PPoEM. In order to evaluate the proposed mechanism, we use 100 prosumers data from AusGrid data~\cite{ausgrid}, and allocate a specified number of prosumers under management of each VPP. This allocation can be varied according to the need, however, for the sake of evaluation and analysis we allocated a prosumers under each VPP according to division as follows: \{VPP 1 = VPP 2 = 3, VPP 3 = 4, VPP 4 = VPP 5 = 5, VPP 6 = VPP 7 = VPP 8 = 10, VPP 9 = 20, and VPP 10 = 30 prosumers\}.
\subsubsection{Mining Leader Determination}
\begin{figure}[t]
\centering
\captionsetup{justification=centering}
\begin{center}
\subfigure[]{
\includegraphics[scale = 0.57]{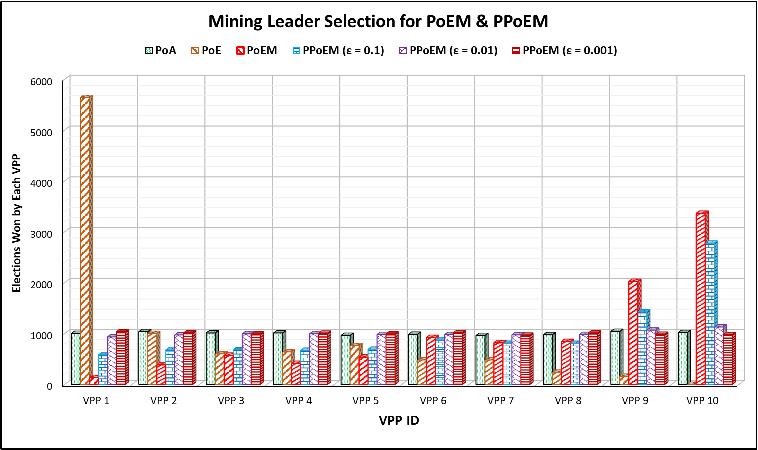}
}
\subfigure[]{
\includegraphics[scale = 0.57]{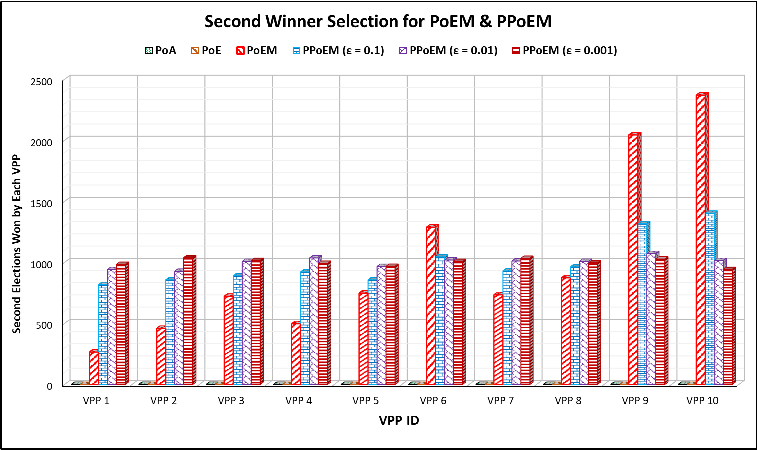}
}
\subfigure[]{
\includegraphics[scale = 0.57]{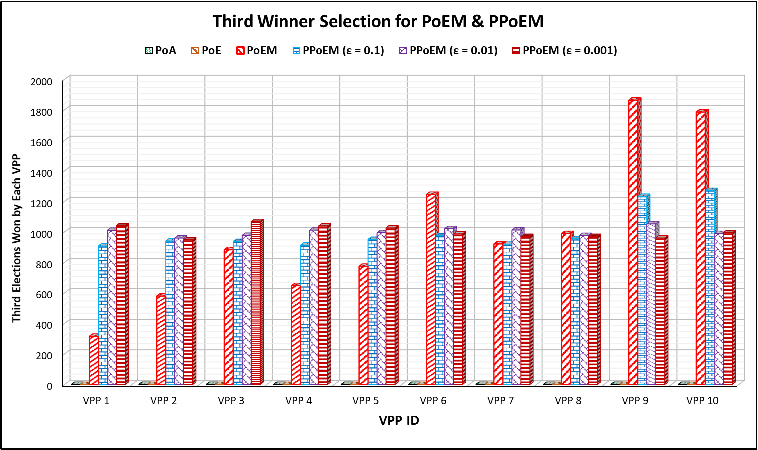}
}

\end{center}
\caption{Performance Comparison of PoEM and PPoEM with PoA~\cite{compare01, compare03, compare04} and PoE~\cite{relnew01}\newline (a) Leader Selection (b) Second Winner Selection (c) Third Winner Selection}
\label{MiningGraph}
\end{figure}

One of the most significant aspects of blockchain consensus mechanism is selection of the winner miner after the election time-out period~\cite{spref15}. In order to do so, our proposed PoEM and PPoEM propose two different strategies. PoEM selects the miner according to the percentage of energy it has traded till the election time-out. However, in PPoEM, the energy traded distribution is further categorized and developed according to the privacy budget. The combined graph for energy mining election outcome is given in Fig. \ref{MiningGraph}. In the graph, each VPP is arranged according to the ascending order of the number of prosumers under it. Similarly, when the number of prosumers increases, the energy traded via that VPP increases. So, it can be determined that VPP 1 is the VPP with least energy trading and VPP 10 being the highest trader among the lot of VPPs. In order to evaluate the mining process, we carry out 10,000 elections on our decentralized blockchain network and in each mining election, the picked energies of VPPs are selected to form a probability distribution for selection of the winning miner.


The mining election is further divided into three steps, in which, first, second, and third winner determination is carried out. In Fig.~\ref{MiningGraph}(a), the winner determination of PoEM, PPoEM is given, it can be \snp{visualized that in both PoEM and PPoEM} ($\varepsilon = 0.1$), VPP 10 wins maximum elections on the basis that it has traded maximum energy, and after that VPP 9 won second highest elections of mining leader. However, the trend equalizes in case of PPoEM ($\varepsilon$ = 0.01 \& 0.001), because of the increase in privacy preservation, all energy mining values are treated approximately equal to others. Furthermore, we compare the work with Proof of Authority (PoA)~\cite{compare01, compare03, compare04} (in which all authority nodes are equal likely to be chosen as miner) and Proof of Energy (PoE)~\cite{relnew01} (in which the prosumer maintaining production-consumption ratio is incentivized). From the experimental evaluation graphs, it can be seen that miner chosen in PoA is equally random and all VPPs are selected equally without any discrimination on basis of energy. Similarly, in PoE~\cite{relnew01}, the prosumer/VPP which maintains production-consumption ratio  near to zero has the highest chances of winning the election and the VPP which trades maximum energy is not incentivized at all. In our case, VPP 1 has the least amount of houses under its control and therefore, it trades minimum energy and has maximum chances of maintaining its ratio, so it wins maximum election via PoE. Contrary to this, the VPP 10, which traded maximum energy won the least elections in PoE because chances of variation are maximum. Therefore, our PoEM incentivizes the VPP trading maximum energy and encourages VPPs to provide incentives to prosumers to trade maximum energy to enhance this trend.\\
Moreover, our proposed PoEM and PPoEM also chooses second and third winner for courtesy reward and provides some proportion of mining fee sum to them in order to encourage maximum VPPs to participate in the mining election by trading maximum energy. Contrary to this, PoA and PoE mechanisms do not provide such features and only provide functionality of leader selection. A comparative analysis of our PoEM and PPoEM with other mining mechanisms is provided in Table.~\ref{tab:compare}. 


\begin{figure}[t]
\centering
\includegraphics[scale=0.63]{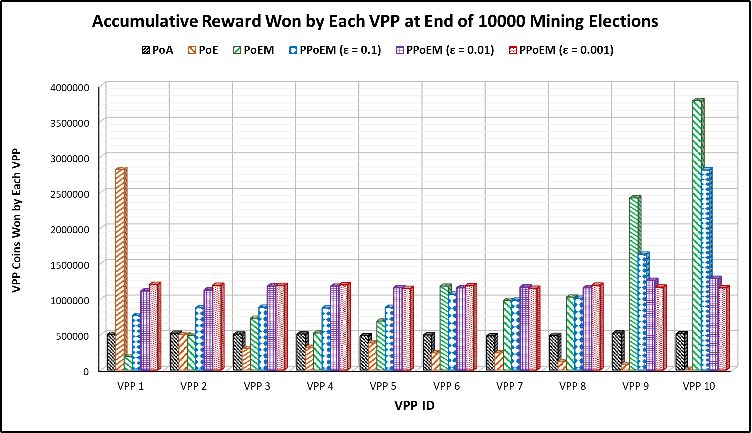}
\caption{Mining Reward/Incentive Comparison of PoEM and PPoEM Mechanism with PoA and PoE after 10,000 Elections.}
\label{fig:incentive}
\end{figure}


\subsubsection{Mining Incentive Determination}
Another important aspect of a mining mechanism is incentivizing the participants. In order to incentivize participants, we provide three factor incentivization in our PoEM and PPoEM mechanisms. Firstly, if a VPP node is selected as winner VPP, it gets incentivized with miner reward (500 VPP coins) and selected percentage of mining sum (this percentage is 70\% in case of PoEM and in PPoEM its selected randomly). Extensive evaluation of accumulated reward won by each VPP after 10,000 mining  elections is given in Fig.~\ref{fig:incentive}. Since VPP No. 10 traded maximum amounts of energy it won the highest accumulated reward in both PoEM and PPoEM strategy as given in the graphical figure. We further compare our PoEM and PPoEM with PoA and PoE mining strategies, which provide miners with the predetermined miner reward and does not provide any extra incentives to second and third miner or to the miner which traded maximum energy.\\ 
\textit{Considering all the discussion, it can be concluded that our proposed PoEM and PPoEM strategies outperform PoA and PoE mechanisms from the perspective of encouraging miners to trade maximum energy.}


\begin{figure}[t]
\centering
\includegraphics[scale=0.75]{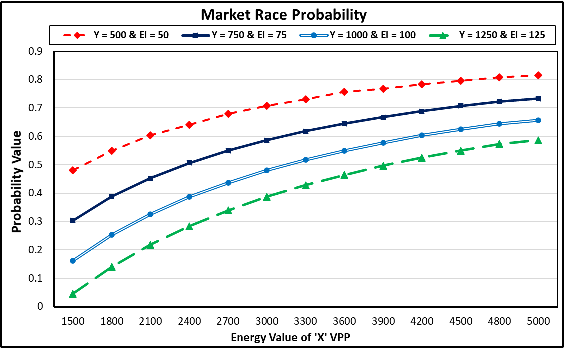}
\caption{Evaluation of Market Race Probability at Multiple Energy Trading Values of VPP 'X' and 'Y'}
\label{marketrace}
\end{figure}



\begin{figure}[t]
\centering
\includegraphics[scale=0.38]{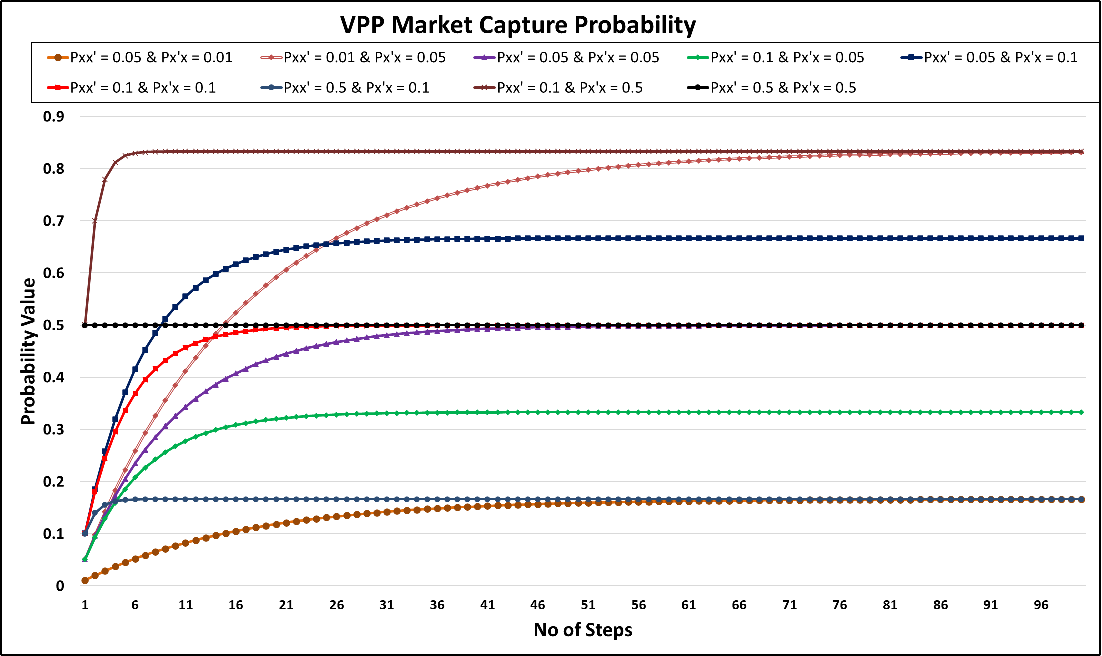}
\caption{Experimental Evaluation of Market Capture Probability for VPP 'X' at Multiple Trading Steps}
\label{markovmarketgraph}
\end{figure}



\begin{figure}[t]
\centering
\includegraphics[scale=0.32]{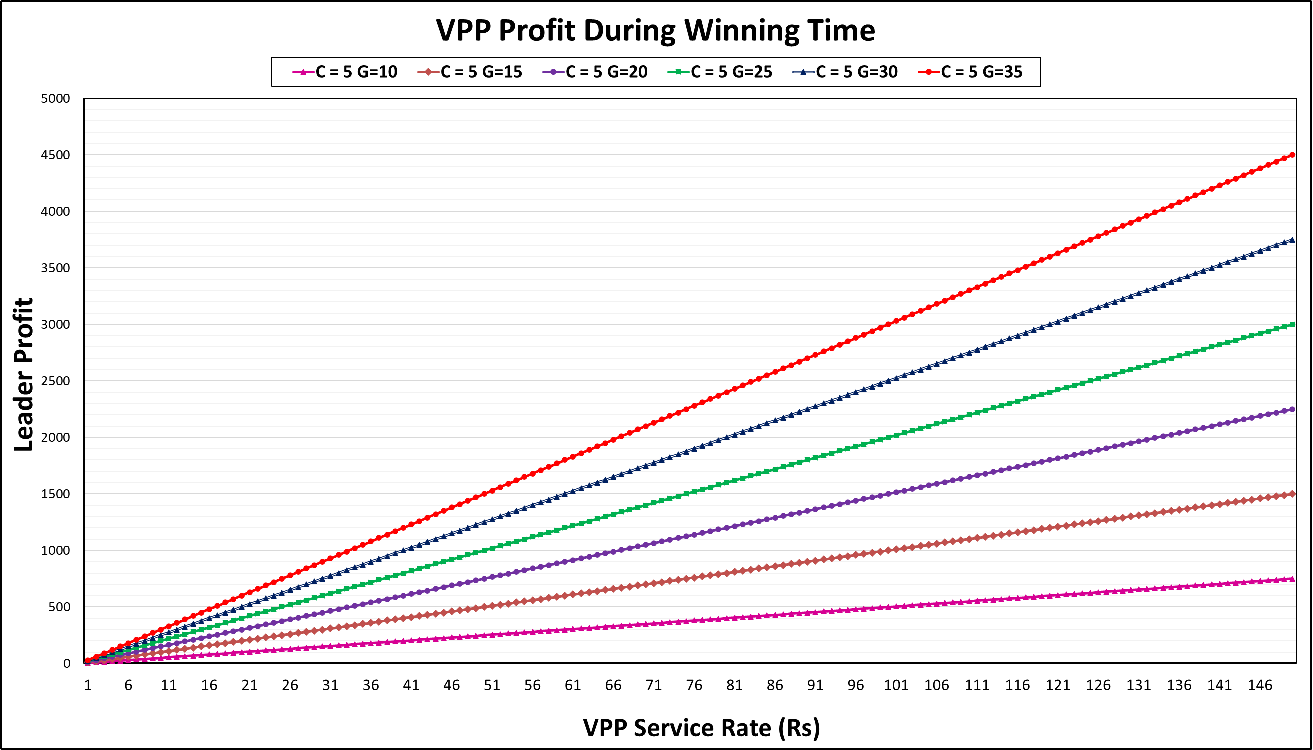}
\caption{Evaluation of Profit Value that a VPP can earn during its Leader Time Period at different Service Rate ($R_s$)}
\label{queuegraph}
\end{figure}


\subsection{Market Capture Probabilities}
The given probabilities help VPPs determine their dominance and \snp{predict that whether a specific VPP can become} one of the leading VPP till the next election time-out.
\subsubsection{Market Race Probability}
Market race probability (evaluated in Fig.~\ref{marketrace}) can be used to determine that at what energy trading value the winning VPP will always remain ahead of other VPPs in order to maximize its chance of winning the mining election. In order to analyse it up, we evaluate the probability value at different energy values of VPP `X’ and `Y’, with VPP `X’ being the leading VPP throughout the time till election time-out. In the graph, `Y’ represents the cumulative energy “$\mathcal{P}_x$” and “$E_i$” represents the last transmitted energy to the mining pool, which is also written as $E_v(i)$ in the above sections. It can be visualized from the graph that when the value of accumulated energy of `Y’ is minimum (e.g., Y = 500), the chances of `X’ VPP leading the election remains maximum. Contrary to this, when the value of `Y’ VPP is increased to 1250, with $E_i$ = 125, the probability of `X’ leading till the election time-out is nearly equal to zero when it has only traded 1500Wh of energy. However, this probability increases with the increase in energy traded value of `X’.
\subsubsection{Market Capture Steady State Probability}
This probability is used to determine the last state of the Markovian Market model, in which each VPP wants to find the final distribution of microgrids at their end in case of transition of participants between these VPPs. We evaluate this for a specific VPP `X', considering the factor that each VPP will be interested in finding out its final share of the market. To evaluate, we consider 9 different transition probability values for Equation~ \ref{ProbT02}, and evaluate it in the Fig. \ref{markovmarketgraph}. From the figure, it can be visualized that when the transition probability from VPP X to other VPPs X$^\prime$ is equal, then the market reaches its stable state in fewer steps as compared to any other probability. This graph can be used by VPPs to visualize how much market they can capture till the next election time-out and what will be the steady state participant distribution for VPPs. 

\subsection{VPP Profit During Leading Time}
A significant parameter that VPPs are usually interested in is the profit they can earn during their leading time. For example, if a VPP wins a mining election, then the next step for it is to validate the transactions and add them in the validated side of the mining pool. As a reward of this validation, VPP gets a percentage of transaction fee, which is directly linked with the profit a VPP will be making during its leading time. It is important to consider that VPPs have limited transaction limits and can also validate the transaction at a specific service rate. Moreover, the running cost of the system is also considered while calculating this profit as given in Eq.~\ref{Que01}. We calculate the prospective profit of leader VPP at different service rate values and provide the results in Fig. \ref{queuegraph}. In the given figure, multiple lines show the calculated profit at different gain (profit per transaction) values. For example, the profit remains minimum when the gain value is 10, however, the profit rises to a maximum peak when the gain value is increased to 35. These values can be used by VPPs to determine a prospective amount of profit they can earn during their leading time with respect to various service rates.\\
\textit{After carefully analysing all experimental results given in graphs, we believe that VPT is the most suitable energy trading model for any decentralized VPP application.}

\section{Conclusion}

In this paper, we work over providing a decentralized demand response enhancing strategy along with proposing \snp{an energy oriented consensus miner selection mechanism named as proof of energy market (PoEM). We further extend this mechanism and integrate differential privacy to protect the privacy of participating DERs, buyers, and VPPs, and name it as private proof of energy market (PPoEM).} Afterwards, we carry out detailed theoretical analysis for differential privacy, security, complexity, and various probabilities such as market race, market stability, prospective profit, which can be used to determine and predict the behaviour and functioning of a complete model. \snp{Overall, we propose a virtual private trading (VPT) model, via which VPPs, DERs, and energy buyers can form a complete system and carry} out energy trading in the most efficient manner. The performance evaluation of the VPT model shows that our proposed model is one of the most suitable choices for a decentralized blockchain based VPP network as compared to other state-of-the-art works.

\appendix


\hspace{10mm}\textit{\textbf{Proof of Theorem~\ref{difPriv01}:}} Consider $W_{bid}$ \& $W_{bid}^\prime$ $\in N^{|X|}$ in a way that $||W_{bid} - W_{bid}^\prime|| \leq $ 1. The arbitrary length of string up to $`l'$ for $W_{bid}$ \& $W_{bid}^\prime$ will be $W_b$ = $\{ W_{b_1}, W_{b_2}, . . . W_{b_l}\}$. Provided, Lap($W_{bid}, F_i, \varepsilon_1$) and Lap($W_{bid}^\prime, F_i, \varepsilon_1$) have probability functions $P_{W_{bid}}$ and $P_{W_{bid}^\prime}$ accordingly. The values can be compared as follows:

\begin{dmath}
\frac{P_{W_{bid}} \left[W_{b} = \{ W_{b_1}, W_{b_2}, . . . W_{b_l}\}\right]}
{P_{W_{bid}^\prime}\left[W_{b} = \{ W_{b_1}, W_{b_2}, . . . W_{b_l}\}\right]} = 
\prod_{k=1}^{l} 
\frac{\exp\left(- \frac{\varepsilon_1 |F_i(W_{bid})_k - W_{b_k}|}{\Delta F_i}\right)}
{\exp\left(- \frac{\varepsilon_1 |F_i(W_{bid}^\prime)_k - W_{b_k}|}{\Delta F_i}\right)}
\end{dmath}

The above equation can be rounded to $\leq \exp(\varepsilon_1)$ considering the Theorem 3.6 in~\cite{dpbook}. Thus, it can be concluded that Laplace mechanism in price selection of PPoEM is $\varepsilon_1$-differentially private.



\hspace{10mm}\textit{\textbf{Proof of Theorem~\ref{difPriv02}:}} Consider $O_p$ is the output range for exponential price selection for two neighboring inputs $P_v$ and $P_v^\prime$. The evaluation can be written as:

\begin{dmath}
\label{EqnTheorem01}
\small
\frac{P_w(F(P_v, q_1, O_p) = o_p)}{P_w(F(P_v^\prime, q_1, O_p) = o_p)} =
\frac{ \frac{\exp (\frac{\varepsilon_2 . q_1(P_v, o_p)}{2 \Delta q_1})}{\sum\limits_{{o_p}^\prime \in O_p} \exp (\frac{\varepsilon_2 . q_1(P_v, {o_p}^\prime)}{2 \Delta q_1})}}
{\frac{\exp (\frac{\varepsilon_2 . q_1(P_v^\prime, o_p)}{2 \Delta q_1})}{\sum\limits_{{o_p}^\prime \in O_p} \exp (\frac{\varepsilon_2 . q_1(P_v^\prime, {o_p}^\prime)}{2 \Delta q_1})}}
\end{dmath}

The privacy budget used in above equation remains $\varepsilon_2$ irrespective of input $P_v$ and $P_v^\prime$. So, Eqn.~\ref{EqnTheorem01} can be rounded to $\exp(\varepsilon_2)$ by using Theorem 3.10 in~\cite{dpbook}. Therefore, it can be concluded that exponential price selection follows $\varepsilon_2$-differential privacy.\\
Similarly, the exponential miner selection of PPoEM have fixed $\varepsilon_3$ privacy budget and follows the same principle of Eqn.~\ref{EqnTheorem01}. Therefore, considering the factors it can be concluded that it obeys $\varepsilon_3$-differential privacy.



\hspace{10mm}\textit{\textbf{Proof of Theorem~\ref{DPComplete}:}} In the PPoEM algorithm, sequential privacy of Laplace and Exponential have been applied step by step in data with privacy parameters $\varepsilon_1$ and $\varepsilon_2$ respectively. Then, according to Lemma~\ref{lemmalabel01}, if sequential privacy steps are performed on same data and it satisfy $\varepsilon_i$-differential privacy on individual privacy level, then it also satisfy a collective $\left(\sum_i \varepsilon_i \right)$-differential privacy. This phenomenon in our PPoEM auction is written as ($\varepsilon_1$ + $\varepsilon_2$)-differential privacy. Since, only two privacy parameters $\varepsilon_1$ \& $\varepsilon_2$ participate in the auction. So, according to the definition $\left( \sum_i \varepsilon_i = \varepsilon_1 + \varepsilon_2 \right)$, the above statement can be generalized as $\varepsilon$-differential privacy. Thus, the given theorem is proved.



\hspace{10mm}\textit{\textbf{Proof of Theorem~\ref{securityanalysis}:}}\\
\textbf{\snp{Wallet Security:}}
\snp{The wallet accounts in our VPT model are protected by certificates and security keys via blockchain cryptographic encryption. So, no one can open or steal the cryptographically protected wallet in our VPT model without the matching certificates and keys.}\\
\textbf{\snp{Transaction Authenticity:}}
\snp{The data in our VPT model is verified at multiple steps in order to maintain its authenticity. Firstly, the winning miner verifies each transaction value individually and adds that to the mining pool during its winning time. After that, the winning miner picks all verified transactions and from the block using SHA-256 hashing. The formed block is then sent to all other nodes for validation, and in case if there is any fraudulent activity, then the specific VPP rejects the block and launches a request for reconsideration of the block. In this way, all the data inside the blocks of our VPT model is ensured to be authentic.}\\
\textbf{\snp{Block Confidentiality}}
In order to protect confidentiality, we integrate two aspects. Firstly, our VPT energy trading runs over a permissioned blockchain network in which only approved nodes can join and view the blockchain ledger. Secondly, we integrate differential privacy as a mean to protect transaction and mining data from internal intruders~\cite{spref26}.\\
\textbf{\snp{Block Integrity}}
 In our permissioned blockchain network, each block integrity is maintained using SHA-256 cryptographic hashing algorithm, which guarantees that all the data inside the block is tamper-proof and is in its original form~\cite{spref28}. Furthermore, in our VPT mechanism, only VPP has the right to mine blocks in the network, which ensures that only approved data will get mined. And in case if any VPP behaves maliciously, then it can be penalized because of its actions.\\
\textbf{\snp{Data Availability}}
Since we are running a system that supports distributed decentralized ledger at each participating node, therefore, the data is always available for viewing on that ledger. Any participating nodes can download the ledger in order to analyse its trading values in order to make sure that the data is tamper-proof.\\
\textbf{\snp{Resilience to Over-provisioning:}}
\snp{Since our proposed model works over permissioned blockchain, therefore, the risk of malicious activity from energy trading nodes is minimal because malicious nodes can be penalized afterwards. E.g., firstly, due to smart metering protocol, none of the energy trading nodes can show/trade more energy than the stored energy. Secondly, in case if some node behaves maliciously, then it can be penalized by giving him a penalty of a specific VPP coin. In worst case scenarios, the membership of a node to join the VPT network can also be cancelled and that node cannot re-join the network. }\\
\textbf{\snp{Fork Resolution:}}
\snp{Formation of forks in the blockchain network have become a worrisome issue due to recent development of blockchain models, because modern blockchains have more computational power as compared to previous blockchain, so they can solve puzzle a bit quicker comparatively~\cite{usenixref01}. However, our proposed VPT model is not vulnerable to the issue of fork formation, because in our model, a block is only generated by the winning miner selected via PoEM or PPoEM mechanism. Therefore, there is only one block per hour and the chances of fork formation are near to zero. }

\hspace{10mm}\textit{\textbf{Proof of Theorem~\ref{attackanalysis}:}} \snp{We develop an energy trading model which is resilient to certain security and privacy attacks due to its private and secure nature. In this section, we discuss these attacks and their resilience in detail.}\\
\textbf{\snp{Sybil Attacks:}}
\snp{In order to enhance trustworthiness in the network, it is important to ensure that none of the participants can make malicious IDs to carry out a sybil attack. Our proposed VPT model is secure from such attacks because in order to join the proposed blockchain network, one must first build a profile and then get approval from authoritative nodes. The approval is given after careful analysis of the profile of the participant, and since an important factor in the profile building is smart meter ID, which is always unique as each smart meter can only be used to create a single account. Therefore, these measures in our VPT model eradicate the possibility of sybil attack in the network.}\\
\textbf{\snp{Inference Attacks}}
\snp{Since our VPT mechanism works over blockchain functionality in which every node has a copy of ledger, therefore, the standard blockchain is vulnerable to inference attack. However, in order to overcome this vulnerability, we integrate the notion of differential privacy in our auction and miner winner selection. Considering the strong theoretical privacy guarantees of our PPoEM proved in Section~\ref{DPAnalysis}, it can be claimed that our model provides strong resilience to inference attacks.}



\hspace{10mm}\textit{\textbf{Proof of Theorem~ \ref{ProbT01} :}} We divide the energy trading time-out till every miner election into equal reporting slots $i$, at which every VPP reports its traded energy to the mining pool for the sake of record. However, it is important to note that the functionality of reading mining pools for formation of blocks is only available at the time of mining election. Let $\mathscr{S}_x$ be the energy traded by $x$ VPP during trading hour and $\mathscr{S}_{y}$ be the energy traded by second highest VPP at that particular $i^{th}$ time during trading period, which means $\mathscr{S}_{x} = \sum_i E_{v_x}(i)$ and $\mathscr{S}_{y} = \max \left[\underset{i}{D_{vpp}} \nexists E_v(i)\right]$.\\
Let $\mathcal{P}_{\mathscr{S}_x, \mathscr{S}_y}$ describe the required probability of $x$ VPP being ahead of average. Let $x$ traded the last reported energy deal at $i^{th}$ time, we did it as $E_v(i)$. The desired probability is as follows:

{\small
\begin{dmath}
\label{Teqn02}
\mathcal{P}_{\mathscr{S}_{x}, \mathscr{S}_{y}} = \left[\mathcal{P} \{x~always~ahead~|~x~traded~last~energy\} * \newline \frac{\mathscr{S}_{x}}{\mathscr{S}_{x}+\mathscr{S}_{y}}\right] + \left[\mathcal{P} \{x~always~ahead~| ~y~traded~last \newline ~energy\} *  \frac{\mathscr{S}_{y}}{\mathscr{S}_{x}+\mathscr{S}_{y}}\right]
\end{dmath}
}

Given that $x$ traded the last energy, we can visualize that probability value of $x$ being ahead than average is same as that of $\mathscr{S}_{x} – E_v(i)$. Similarly, when $y$ traded the last energy, then reduction of probability will be according to $\mathscr{S}_{y} - E_v(i)$. Then the Eqn.~\ref{Teqn02} can be written as:

\begin{dmath}
\label{Teqn03}
\mathcal{P}_{\mathscr{S}_{x}, \mathscr{S}_{y}} = \left[\frac{\mathscr{S}_{x}}{\mathscr{S}_{x}+\mathscr{S}_{y}} * \mathcal{P}_{\left(\left[\mathscr{S}_{x} - E_v(i) \right],\mathscr{S}_{y}\right)}\right] + \left[\frac{\mathscr{S}_{y}}{\mathscr{S}_{x}+\mathscr{S}_{y}} * \mathcal{P}_{\left(\mathscr{S}_{x},\left[\mathscr{S}_{y} - E_v(i)\right]\right)}\right]
\end{dmath}

As it is true  that when only a single trade $E_v(i)$ happened, then according to induction hypothesis $\mathscr{S}_x + \mathscr{S}_u = 1$ and $\mathscr{S}_x + \mathscr{S}_u = k$ to $\mathscr{S}_x + \mathscr{S}_u = k+1$, Eqn.~\ref{Teqn03} can be written as:

\begin{dmath}
\mathcal{P}_{\mathscr{S}_{x}, \mathscr{S}_{y}} = \left[\frac{\mathscr{S}_{x}}{\mathscr{S}_{x}+\mathscr{S}_{y}} * \frac{\mathscr{S}_{x} – E_v(i) - \mathscr{S}_y}{\mathscr{S}_{x} – E_v(i) + \mathscr{S}_y }\right] + \left[\frac{\mathscr{S}_{y}}{\mathscr{S}_{x}+\mathscr{S}_{y}} * \frac{\mathscr{S}_{x} - \mathscr{S}_y + E_v(i) }{\mathscr{S}_{x}  + \mathscr{S}_y - E_v(i) }\right]
\end{dmath}

\begin{align*}
 = \frac{\mathscr{S}_{x^2}-\mathscr{S}_x E_v(i) - \mathscr{S}_x \mathscr{S}_y + \mathscr{S}_x \mathscr{S}_y - \mathscr{S}_{y^2} +\mathscr{S}_y E_v(i) }{(\mathscr{S}_x + \mathscr{S}_y)( \mathscr{S}_x + \mathscr{S}_y – E_v(i))}
\end{align*}

\begin{align*}
 = \frac{\mathscr{S}_{x^2} - \mathscr{S}_{y^2} - \mathscr{S}_{x} E_v(i) - \mathscr{S}_{y}E_v(i)}
{\mathscr{S}_{x}E_v(i) - \mathscr{S}_{y}E_v(i) + \mathscr{S}_{x^2} + \mathscr{S}_{y^2} + 2\mathscr{S}_{x}\mathscr{S}_{y}}
\end{align*}

The final equation will be written as:

\begin{equation}
\label{Teqn04}
\mathcal{P}_{\mathscr{S}_{x}, \mathscr{S}_{y}} = \frac{\mathscr{S}_{x}(\mathscr{S}_{x}-E_v(i)) - \mathscr{S}_{y}(\mathscr{S}_{y} + E_v(i))    }{\mathscr{S}_{x}(\mathscr{S}_{x} - E_v(i)) + \mathscr{S}_{y}(\mathscr{S}_{y} - E_v(i)) + 2\mathscr{S}_{x}\mathscr{S}_{y}}
\end{equation}

Hence the theorem is proved.



\hspace{10mm}\textit{\textbf{Proof of Theorem~\ref{ProbT02}:}} The transition probabilities for VPP $x$ and VPP $x’$ can be derived from the state diagram given in Fig.~\ref{fig:MarkovGraph}(b). The state matrix will be as follows:

\[
\mathcal{P}_r = 
\begin{bmatrix}
1 - \mathcal{P}_{xx’} & \mathcal{P}_{xx’}\\
\mathcal{P}_{x’x} & 1 - \mathcal{P}_{x’x}
\end{bmatrix}
\]

Current VPP probability distribution ($\Upsilon_{_d} = [C_x~~~C_{x’}]$) demonstrate the microgrids served by VPP $x$ and VPPs $x'$. So, the steady state distribution can be calculated as follows:

\begin{align*}
\Upsilon_{_d} * \mathcal{P}_r^{^m} = \Upsilon_{_d} 
\end{align*}

\begin{dmath}
\label{matrixeqn01}
\begin{bmatrix}
C_x & C_{x’}
\end{bmatrix}
\begin{bmatrix}
1 - \mathcal{P}_{xx’} & \mathcal{P}_{xx’}\\
\mathcal{P}_{x’x} & 1 - \mathcal{P}_{x’x}
\end{bmatrix}
= 
\begin{bmatrix}
C_x & C_{x’}
\end{bmatrix}
\end{dmath}

\begin{align*}
{\Upsilon_{_d}^{^m}} * \mathcal{P}_r = {\Upsilon_{_d}^{^m}} 
\end{align*}

\begin{align*}
{\Upsilon_{_d}^{^m}} \mathcal{P}_r - {\Upsilon_{_d}^{^m}} = 0 
\end{align*}

\begin{equation}
\label{matrixeqn02}
{\Upsilon_{_d}^{^m}}(\mathcal{P}_r - I) = 0 
\end{equation}

Furthermore, Eqn.~\ref{matrixeqn01} and Eqn.~\ref{matrixeqn02} can be combined as follows:

\begin{dmath}
\begin{bmatrix}
C_x & C_{x’}
\end{bmatrix}
\begin{bmatrix}
\begin{pmatrix}
1 - \mathcal{P}_{xx’} & \mathcal{P}_{xx’}\\
\mathcal{P}_{x’x} & 1 - \mathcal{P}_{x’x}
\end{pmatrix}
- 
\begin{pmatrix}
1 & 0\\
0 & 1
\end{pmatrix}
\end{bmatrix}
= 0
\end{dmath}

\begin{dmath}
\label{matrixeqn03}
\begin{bmatrix}
C_x & C_{x’}
\end{bmatrix}
\begin{bmatrix}
\begin{pmatrix}
\mathcal{P}_{xx’} & \mathcal{P}_{xx’}\\
\mathcal{P}_{x’x} & - \mathcal{P}_{x’x}
\end{pmatrix}
\end{bmatrix}
= 0
\end{dmath}

\begin{dmath}
\begin{bmatrix}
-C_x \mathcal{P}_{xx'} + C_{x'}\mathcal{P}_{x'x} & C_x \mathcal{P}_{xx'} - C_{x'}\mathcal{P}_{x'x}
\end{bmatrix}
= 
\begin{bmatrix}
0 & 0
\end{bmatrix}
\end{dmath}

We can now derive two equations accordingly:

\begin{equation}
\label{matrixeqn04}
-C_x \mathcal{P}_{xx'} + C_{x'}\mathcal{P}_{x'x} = 0
\end{equation}

\begin{equation}
C_x \mathcal{P}_{xx'} - C_{x'}\mathcal{P}_{x'x} = 0
\end{equation}

According to Markov probability distribution rules for Fig.~\ref{fig:MarkovGraph}(a), the sum of all probabilities should be 1, which is derived as follows:

\begin{equation}
\sum_{i=1}^{V_N} C_{(i)} = C_x + C_{y1} + C_{y2} +. . . . + C_{V_N} = 1
\end{equation}
The above equation can further be reduced for Fig.~\ref{fig:MarkovGraph}(b) as follows:

\begin{equation}
\label{matrixeqn05}
C_{(i)} = C_x + C_{x'} = 1
\end{equation}

Eqn~\ref{matrixeqn04} can be reduced using Eqn.~\ref{matrixeqn05} as follows:

\begin{equation}
-C_x \mathcal{P}_{xx'} + C_{x'}\mathcal{P}_{x'x} + \mathcal{P}_{xx'}C_x + \mathcal{P}_{xx'}C_{x'} = \mathcal{P}_{xx'}
\end{equation}

\[
C_{x'}\left(\mathcal{P}_{x'x} + \mathcal{P}_{xx'}\right) = 1
\]

\begin{equation}
\label{matrixeqn06}
C_{x'} = \frac{\mathcal{P}_{xx'}}{\mathcal{P}_{x'x} + \mathcal{P}_{xx'}}
\end{equation}

The value of Eq.~\ref{matrixeqn05} can be substituted in Eq.~\ref{matrixeqn06} as follows:

\[
C_{x} + \frac{\mathcal{P}_{xx'}}{\mathcal{P}_{x'x}+\mathcal{P}_{xx'}} = 1
\]

\[
C_{x} = 1 - \frac{\mathcal{P}_{xx'}}{\mathcal{P}_{x'x}+\mathcal{P}_{xx'}}
\]

\[
C_x = \frac{\mathcal{P}_{x'x}+P_{xx'}-\mathcal{P}_{xx'}}{\mathcal{P}_{x'x}+\mathcal{P}_{xx'}}
\]

Thus final equation can be derived as follows:
\begin{equation}
C_x = \frac{\mathcal{P}_{x'x}}{\mathcal{P}_{x'x}+\mathcal{P}_{xx'}}
\end{equation}
From results of \textit{Proof} the required theorem is proved.


\hspace{10mm}\textit{\textbf{Proof of Theorem~\ref{ProbT05}:}} Let $\pi_{\alpha}$ (where $\alpha = $ 1, 2, ..... $n$) denote proportions of long-run for VPPs. Therefore, $j\in T_1$, $k \in T_2$ \& $l \in  T_3$ at which VPP enters state of $T_1$ after fulfilling initial condition according to Table.~\ref{tab:STP01}.\\
The probability value $P_{E_k}(x) \geq 20\%$ means that:

\begin{dmath}
{P_{E_k}(x) \geq 20\% = \left[\frac{\mathscr{S}_x}{\mathscr{S}_{sum}} * 100\right] \geq 20\%}  = {\left[\frac{\sum_i^n E_{V_x}(i)}{\sum_j^{V_N} \sum_i^n E_{V_j}(i)}*100\right] \geq 20\%}
\end{dmath}

The above probability can further be defined as follows:
\vspace{-0.4em}
\[
\underline{For~j~to~k:}~Rate~of~entering~k~from~j~=~\pi_i P_{jk}
\]
\vspace{-1.1em}
\[
Rate~of~entering~k~from~T_1~=~\sum_{j\in T_1} \pi_j P_{jk}
\]
\vspace{-1.1em}
\[
\underline{For~j~to~l:}~Rate~of~entering~l~from~j~=~\pi_j P_{jl}
\]
\vspace{-1.1em}
\[
Rate~of~entering~l~from~T_1~=~\sum_{j\in T_1} \pi_j P_{jl}
\]

The combined probability can be derived as: 

{\small
\begin{dmath}
Accumulative~rate~of~entering~l~\&~k~from~T_1 = \sum_{j\in T_1} \pi_j P_{jk} + \sum_{j\in T_1} \pi_j P_{jl}
\end{dmath}
}

\[
= \sum_{j\in T_1} \pi_j (P_{jk}+P_{il})
\]

Similarly, rate at which low winning VPP state occur which is also known as occurrence~rate~of~low~winning~state (ORLS) is derived as follows:

\begin{dmath}
\label{ORLS01}
ORLS = \sum_{l\in T_3} \sum_{k\in T_2}\left(\sum_{j\in T_1} \pi_j (P_{jk}+ P_{jl}\right)
\end{dmath}

We further denote high probability winning average time and low probability time as $\vec{WT}$ and $\vec{LT}$ respectively. Moreover, a single transition occur when a VPP transit from high state $\vec{WT}$ to low state $\vec{LT}$ in $(\vec{WT}+\vec{LT})$ unit as average. So, it can further be modelled as:

\begin{dmath}
\label{ORLS02}
ORLS = \frac{1}{\vec{WT}+\vec{LT}}
\end{dmath}

So, Eq.~\ref{ORLS01} and Eq.~\ref{ORLS02} can be equated as:

\begin{dmath}
\label{ORLS03}
\frac{1}{\vec{WT}+\vec{LT}} = \sum_{l\in T_3} \sum_{k\in T_2}\left(\sum_{j\in T_1} \pi_j (P_{jk}+ P_{jl}\right)
\end{dmath}

The probability via which a VPP is in winning state is $\sum_{j\in T_1}\pi_j$. Although, the average time at which VPP remains in winning state ($\vec{WT}$) is averaged with respect to total duration ($\vec{WT} + \vec{LT}$). Therefore, it is formally represented as:
\vspace{-1.2em}
\[
Winning~Time~Relative~Proportion= \frac{\vec{WT}}{\vec{WT} + \vec{LT}}
\]

This can further be formulated to:

\[
\frac{\vec{WT}}{\vec{WT} + \vec{LT}} = \sum_{j\in T_1} \pi_j
\]

\[
\vec{WT} = \sum_{j\in T_1} \pi_j (\vec{WT}+\vec{LT})
\]

\begin{dmath}
\label{ORLS04}
\vec{WT} = \frac{\sum_{j\in T_1} (\pi_j)}{\frac{1}{\vec{WT} + \vec{LT}}}
\end{dmath}

The value of Eq.~\ref{ORLS03} can be substituted in Eq.~\ref{ORLS04} to get the final results to prove the theorem as follows:

\begin{dmath}
\vec{WT} = \frac{\sum_{j\in T_1} (\pi_j)}{\sum_{l\in T_3}\sum_{k\in T_2}\sum_{j \in T_1} \pi_j (P_{jk} + P_{jl})}
\end{dmath}

\hspace{10mm}\textit{\textbf{Proof of Theorem~\ref{LeadTheorem}:}} Let $R_A$ be the rate of arrival of transactions from a follower VPP and $R_s$ be the rate of service of leader VPP. Similarly, at a given time, the selected VPP can deal with a maximum $T_L$ number of transactions. Moreover, the amount that a VPP spends during its validation period will be determined via  $C R_s$. Taking into consideration the transaction arrival, one can compute leader profit via $R_a (1- P_{T_L})M$. To calculate $P_{T_L}$, the initial single VPP queueing model assumptions is used by considering arrival and service rate as follows:

\begin{equation}
\label{Que02}
P_{T_L} = \left(\frac{R_A}{R_s}\right)^T_L . P_o ~~ T= \{1,2,. . . T_L\}
\end{equation}

The value of $P_O$ can be solved from above equation as follows:
\[
1 = \sum_{T = 0}^{T_l}\left(\frac{R_A}{R_s}\right)^T P_o
\]

\begin{align}
P_o = \frac{1-\frac{R_A}{R_s}}{1-\left(1-\frac{R_A}{R_s}\right)^{T_L+1}}
\end{align}

The value of above equation can be substituted in basic formula as follows: 

\begin{align}
\label{Que07}
P_{T_L} = \frac{\left(\frac{R_A}{R_s}\right)^{T_L} \left(1-\frac{R_A}{R_s}\right)}{1-\left(\frac{R_A}{R_s}\right)^{T_L +1}}
\end{align}

The above equation can further be substituted in the basic profit formula as:

\begin{align}
\scriptsize
\label{Que08}
T_P = R_A . M \left[ \frac{1 - \left(\frac{R_A}{R_s}\right)^{T_L} \left(1-\frac{R_A}{R_s}\right)}{ 1-\left(\frac{R_A}{R_s}\right)^{T_L +1}}\right]
 ~-~ C R_s
\end{align}

\[\scriptsize
\label{Que09}
T_P = R_A . M \frac{\left[1 - \left( \frac{R_A}{R_s}  \right)^{T_L+1} -\left(\frac{R_A}{R_s}\right)^{T_L} + \left(\frac{R_A}{R_s}\right)^{T_L+1}\right]}
{ 1-\left(\frac{R_A}{R_s}\right)^{T_L +1}} ~-~ C R_s
\]
So, the final equation can be written as:
\begin{equation}
\label{Que01}
T_p = \frac{R_A M \left[1 - \left(\frac{R_A}{R_s}\right)^{T_L}\right]}{1-\left(\frac{R_A}{R_s}\right)^{T_L+1}} ~–~ C R_s
\end{equation}
Hence, the prospective profit theorem is proved.


\hspace{10mm}\textit{\textbf{Proof of Theorem~\ref{theorem01}:}} The computational complexity of the first part mainly depends upon the complexity of sorting mechanism and double auction buyer and price selection. Line 1 of PoEM algorithm sorts all ‘N’ number of buyers, whose complexity according to big $\mathcal{O}$ notation is~$\mathcal{O}(N\log(N))$. Afterwards the ‘nested for’ loops are started off, whose complexity totally depends upon the number of buyers and sellers. For instance, the complexity of the outer ‘for’ loop by keeping into consideration all assignment operations will be $\mathcal{O}(3*(S+2))$, which is further rounded to $\mathcal{O}(S)$. Similarly, the complexity of the inner ‘for’ loop after rounding to big $\mathcal{O}$ notation is $\mathcal{O} (SN)$. The further statements from Line 4 – 11 have constant time complexity of $\mathcal{O}(1)$, because they are executed only once. However, in the presence of two ‘nested for’ loops, this constant complexity of the statements will depend upon the complexity of loops which is $\mathcal{O}(SN)$. An important thing to notice over here is that the ‘$argmax()$’ function on Line 5 actually depicts the complete auction process of part 1 of PoEM that is also presented in the respective section, this function is just used here hypothetically to demonstrate the process. Moving further to Line 13 \& 14, these are just append functions having constant computational complexity. So, it can be said that the complexity highly depends upon two factors, ‘$N\log(N)$' \& ‘$SN$'. Therefore, upper bound complexity of auction part of PoEM can be written as $\mathcal{O} (max{N\log(N), SN})$.\\
Hence, the theorem is proved.


\hspace{10mm}\textit{\textbf{Proof of Theorem~\ref{theorem02}:}} The second part of the PoEM algorithm mainly consists of computation of transaction and mining fees along with social welfare, which are computationally constant $\mathcal{O}(1)$ command. However, due to the presence of the ‘for’ loop, the computational complexity of this complete process is $\mathcal{O}(W_s)$. $W_s$ is the list of winning sellers from the part 1 of PoEM, whose value can be a maximum number of sellers ‘$S$’. The upper bound of $W_s$ can be according to total number of sellers $S$, which can formally be equated as $\mathcal{O}(W_s) \cong \mathcal{O}(S)$.\\


\hspace{10mm}\textit{\textbf{Proof of Theorem~\ref{theorem03}:}} The third part of PoEM algorithm comprises of selection and rewarding mining for their participation in the mining process. In this process, basic single step statements from Line 25 – 27, and then from Line 40 – 44 have constant time complexity of $\mathcal{O}(1)$. However, the major computational complexity of this part is caused due to summation and loops, firstly, the summation at Line 28 has a complexity of $\mathcal{O}(V_N)$, which depend upon the total number of VPPs participating in the mining process. Afterwards, the ‘for’ loop also has similar computational complexity because all the statements inside the loop are computationally constant and totally depend upon the iterations by loop, which is $\mathcal{O}(V_N)$. The complexity from Line 34 – 36 is dependent upon the random selection algorithm, which has maximum complexity of $\mathcal{O}(n)$ [in our case $\mathcal{O}(V_N)$. The complexity for Line 34 will be $\mathcal{O}(V_N)$, while in Line 35 and Line 36, one VPP is reduced step by step, so the complexity will be $\mathcal{O}(V_{N-1})$ and $\mathcal{O}(V_{N-2})$ respectively. Calculating mining sum at Line 39 also contribute in overall computational complexity of this part having a value of $\mathcal{O}(M_{x_{f}})$, which is equivalent to $\mathcal{O}(Ws_{max})$. So, the complexity of the third part of PoEM algorithm can be combined as $\mathcal{O}(\max\{Vpp_{n} , Ws_{max}\})$. Keeping in view the fact that number of participating sellers is much larger than number of VPPs ($Ws_{max} \gg Vpp_{n}$), the complexity can be simplified as $\mathcal{O}(Ws_{max})$.\\

\bibliographystyle{IEEEtran}

\begin{thebibliography}{10}

\bibitem{relref07}
Ana Baringo, Luis Baringo, and Jos{\'e}~M Arroyo.
\newblock Day-ahead self-scheduling of a virtual power plant in energy and
  reserve electricity markets under uncertainty.
\newblock {\em IEEE Transactions on Power Systems}, 34(3):1881--1894, 2018.

\bibitem{usenixref01}
Canhui Chen, Xu~Chen, Jiangshan Yu, Weigang Wu, and Di~Wu.
\newblock Impact of temporary fork on the evolution of mining pools in
  blockchain networks: An evolutionary game analysis.
\newblock {\em IEEE Transactions on Network Science and Engineering},
  8(1):400--418, 2020.

\bibitem{poem02}
Zhili Chen, Tianjiao Ni, Hong Zhong, Shun Zhang, and Jie Cui.
\newblock Differentially private double spectrum auction with approximate
  social welfare maximization.
\newblock {\em IEEE Transactions on Information Forensics and Security},
  14(11):2805--2818, 2019.

\bibitem{spref10}
Zeyu Ding, Yuxin Wang, Guanhong Wang, Danfeng Zhang, and Daniel Kifer.
\newblock Detecting violations of differential privacy.
\newblock In {\em Proceedings of the ACM SIGSAC Conference on Computer and
  Communications Security}, pages 475--489, 2018.

\bibitem{intref00}
Cynthia Dwork.
\newblock Differential privacy.
\newblock In {\em Automata, Languages and Programming}, pages 1--12, Berlin,
  Heidelberg, 2006. Springer Berlin Heidelberg.

\bibitem{dpbook}
Cynthia Dwork, Aaron Roth, et~al.
\newblock The algorithmic foundations of differential privacy.
\newblock {\em Foundations and Trends in Theoretical Computer Science},
  9(3-4):211--407, 2014.

\bibitem{compare01}
Jacob Eberhardt, Marco Peise, Dong-Ha Kim, and Stefan Tai.
\newblock {Privacy-Preserving Netting in Local Energy Grids}.
\newblock {\em IEEE International Conference on Blockchain and Cryptocurrency},
  2020.

\bibitem{compare04}
EnergyChain.
\newblock {\em {The Energy Web Chain Project Online:
  https://www.energyweb.org/technology/energy-web-chain}}.

\bibitem{analysis02}
Nicholas Etherden, Valeriy Vyatkin, and Math~HJ Bollen.
\newblock {Virtual power plant for grid services using IEC 61850}.
\newblock {\em IEEE Transactions on Industrial Informatics}, 12(1):437--447,
  2015.

\bibitem{spref11}
M.~N. {Faqiry} and S.~{Das}.
\newblock Double auction with hidden user information: Application to energy
  transaction in microgrid.
\newblock {\em IEEE Transactions on Systems, Man, and Cybernetics: Systems},
  49(11):2326--2339, 2019.

\bibitem{compare03}
Magda Foti and Manolis Vavalis.
\newblock Blockchain based uniform price double auctions for energy markets.
\newblock {\em Applied Energy}, 254:113604, 2019.

\bibitem{poem01}
Ummy Habiba and Ekram Hossain.
\newblock Auction mechanisms for virtualization in 5g cellular networks:
  Basics, trends, and open challenges.
\newblock {\em IEEE Communications Surveys \& Tutorials}, 20(3):2264--2293,
  2018.

\bibitem{relnew06}
M.~U. {Hassan}, M.~H. {Rehmani}, and J.~{Chen}.
\newblock Deal: Differentially private auction for blockchain-based microgrids
  energy trading.
\newblock {\em IEEE Transactions on Services Computing}, 13(2):263--275, 2020.

\bibitem{prelref04}
M.~U. {Hassan}, M.~H. {Rehmani}, and J.~{Chen}.
\newblock Differential privacy techniques for cyber physical systems: A survey.
\newblock {\em IEEE Communications Surveys Tutorials}, 22(1):746--789, 2020.

\bibitem{spref09}
Mohammad Jadidbonab, Behnam Mohammadi-Ivatloo, Mousa Marzband, and Pierluigi
  Siano.
\newblock Short-term self-scheduling of virtual energy hub plant within thermal
  energy market.
\newblock {\em IEEE Transactions on industrial electronics}, 68(4):3124--3136,
  2020.

\bibitem{spref23}
Y.~{Jiao}, P.~{Wang}, D.~{Niyato}, and K.~{Suankaewmanee}.
\newblock Auction mechanisms in cloud/fog computing resource allocation for
  public blockchain networks.
\newblock {\em IEEE Transactions on Parallel and Distributed Systems},
  30(9):1975--1989, 2019.

\bibitem{spref05}
Harry Kalodner, Malte M{\"o}ser, Kevin Lee, Steven Goldfeder, Martin Plattner,
  Alishah Chator, and Arvind Narayanan.
\newblock Blocksci: Design and applications of a blockchain analysis platform.
\newblock In {\em 29th USENIX Security Symposium)}, pages 2721--2738, 2020.

\bibitem{spref08}
D.~{Koraki} and K.~{Strunz}.
\newblock Wind and solar power integration in electricity markets and
  distribution networks through service-centric virtual power plants.
\newblock {\em IEEE Transactions on Power Systems}, 33(1):473--485, 2018.

\bibitem{spref17}
D.~{Li}, Q.~{Yang}, W.~{Yu}, D.~{An}, Y.~{Zhang}, and W.~{Zhao}.
\newblock Towards differential privacy-based online double auction for smart
  grid.
\newblock {\em IEEE Transactions on Information Forensics and Security},
  15:971--986, 2020.

\bibitem{relref06}
Peikai Li, Yun Liu, Huanhai Xin, and Xichen Jiang.
\newblock A robust distributed economic dispatch strategy of virtual power
  plant under cyber-attacks.
\newblock {\em IEEE Transactions on Industrial Informatics}, 14(10):4343--4352,
  2018.

\bibitem{spref07}
Z.~{Liang}, Q.~{Alsafasfeh}, T.~{Jin}, H.~{Pourbabak}, and W.~{Su}.
\newblock Risk-constrained optimal energy management for virtual power plants
  considering correlated demand response.
\newblock {\em IEEE Transactions on Smart Grid}, 10(2):1577--1587, 2019.

\bibitem{spref16}
Giulio Malavolta, Pedro Moreno-Sanchez, Clara Schneidewind, Aniket Kate, and
  Matteo Maffei.
\newblock Anonymous multi-hop locks for blockchain scalability and
  interoperability.
\newblock In {\em NDSS}, 2019.

\bibitem{spref27}
M.~{Maruseac} and G.~{Ghinita}.
\newblock Precision-enhanced differentially-private mining of high-confidence
  association rules.
\newblock {\em IEEE Transactions on Dependable and Secure Computing}, pages
  1--1, 2018.

\bibitem{spref28}
Benny Pinkas, Thomas Schneider, and Michael Zohner.
\newblock Scalable private set intersection based on ot extension.
\newblock {\em ACM Transactions on Privacy and Security (TOPS)}, 21(2):1--35,
  2018.

\bibitem{spref12}
N.~{Pourghaderi}, M.~{Fotuhi-Firuzabad}, M.~{Moeini-Aghtaie}, and
  M.~{Kabirifar}.
\newblock Commercial demand response programs in bidding of a technical virtual
  power plant.
\newblock {\em IEEE Transactions on Industrial Informatics}, 14(11):5100--5111,
  2018.

\bibitem{spref24}
Fang-Yu Rao, Jianneng Cao, Elisa Bertino, and Murat Kantarcioglu.
\newblock Hybrid private record linkage: Separating differentially private
  synopses from matching records.
\newblock {\em ACM Transactions on Privacy and Security (TOPS)}, 22(3):1--36,
  2019.

\bibitem{relnew01}
Pierluigi Siano, Giuseppe De~Marco, Alejandro Rolán, and Vincenzo Loia.
\newblock A survey and evaluation of the potentials of distributed ledger
  technology for peer-to-peer transactive energy exchanges in local energy
  markets.
\newblock {\em IEEE Systems Journal}, 13(3):3454--3466, 2019.

\bibitem{ausgrid}
SolarHomes.
\newblock {\em {AusGrid: Solar Home Electricity Data Online:
  [https://www.ausgrid.com.au/Industry/Our-Research/Data-to-share/Solar-home-electricity-data]
  }}.

\bibitem{spref14}
Zhou Su, Yuntao Wang, Qichao Xu, and Ning Zhang.
\newblock Lvbs: Lightweight vehicular blockchain for secure data sharing in
  disaster rescue.
\newblock {\em IEEE Transactions on Dependable and Secure Computing},
  19(1):19--32, 2022.

\bibitem{spref15}
Pawel Szalachowski, Daniel Reijsbergen, Ivan Homoliak, and Siwei Sun.
\newblock Strongchain: Transparent and collaborative proof-of-work consensus.
\newblock In {\em 28th USENIX Security Symposium)}, pages 819--836, 2019.

\bibitem{keyref01}
K.~{Wang}, J.~{Yu}, Y.~{Yu}, Y.~{Qian}, D.~{Zeng}, S.~{Guo}, Y.~{Xiang}, and
  J.~{Wu}.
\newblock A survey on energy internet: Architecture, approach, and emerging
  technologies.
\newblock {\em IEEE Systems Journal}, 12(3):2403--2416, 2018.

\bibitem{spref21}
Y.~{Wang}, Z.~{Cai}, Z.~{Zhan}, Y.~{Gong}, and X.~{Tong}.
\newblock An optimization and auction-based incentive mechanism to maximize
  social welfare for mobile crowdsourcing.
\newblock {\em IEEE Transactions on Computational Social Systems},
  6(3):414--429, 2019.

\bibitem{spref26}
M.~{Ye} and A.~{Barg}.
\newblock Optimal schemes for discrete distribution estimation under locally
  differential privacy.
\newblock {\em IEEE Transactions on Information Theory}, 64(8):5662--5676,
  2018.

\bibitem{spref02}
Z.~{Yi}, Y.~{Xu}, W.~{Gu}, and W.~{Wu}.
\newblock A multi-time-scale economic scheduling strategy for virtual power
  plant based on deferrable loads aggregation and disaggregation.
\newblock {\em IEEE Transactions on Sustainable Energy}, 11(3):1332--1346,
  2020.

\bibitem{spref03}
H.~{Yu}, I.~{Nikolić}, R.~{Hou}, and P.~{Saxena}.
\newblock Ohie: Blockchain scaling made simple.
\newblock In {\em IEEE Symposium on Security and Privacy (SP)}, pages 90--105,
  2020.

\bibitem{spref04}
R.~{Zhang} and B.~{Preneel}.
\newblock Lay down the common metrics: Evaluating proof-of-work consensus
  protocols' security.
\newblock In {\em IEEE Symposium on Security and Privacy (SP)}, pages 175--192,
  2019.

\bibitem{spref06}
Runfan Zhang and Branislav Hredzak.
\newblock Distributed dynamic clustering algorithm for formation of
  heterogeneous virtual power plants based on power requirements.
\newblock {\em IEEE Transactions on Smart Grid}, 12(1):192--204, 2020.

\bibitem{spref19}
Haotian Zhao, Bin Wang, Zhaoguang Pan, Hongbin Sun, Qinglai Guo, and Yixun Xue.
\newblock Aggregating additional flexibility from quick-start devices for
  multi-energy virtual power plants.
\newblock {\em IEEE Transactions on Sustainable Energy}, 12(1):646--658, 2020.

\end{thebibliography}

\begin{IEEEbiography}[{\includegraphics[width=1in,height=1.25in,clip,keepaspectratio]{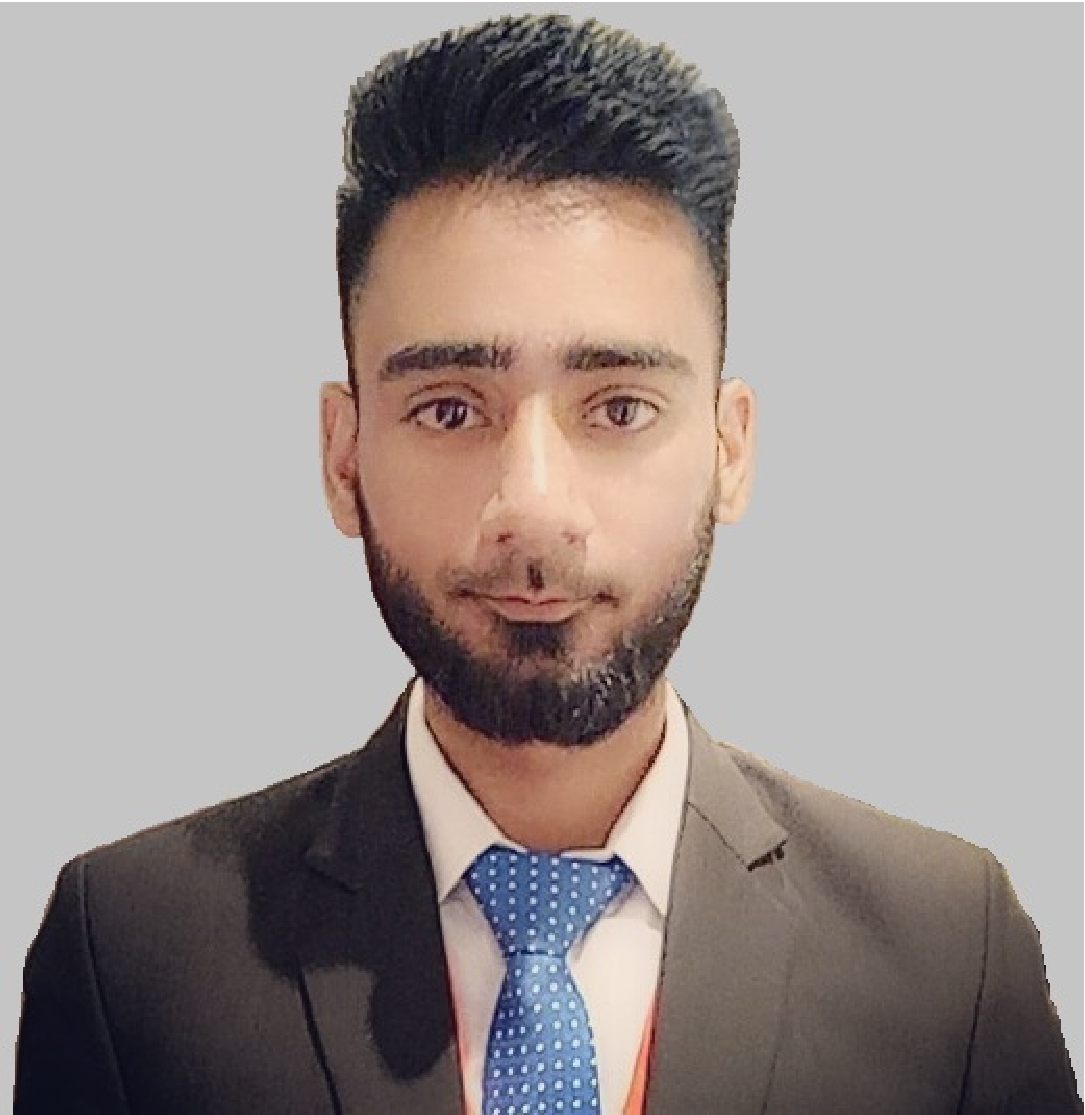}}]{Dr. Muneeb Ul Hassan}

received his PhD degree from Swinburne University of Technology, Australia. He received his Bachelor degree in Electrical Engineering from COMSATS Institute of Information Technology, Wah Cantt, Pakistan, in 2017. He received Gold Medal in Bachelor degree for being topper of Electrical Engineering Department. Currently, he is working as a Research Fellow/ Postdoctoral Researcher at Swinburne University of Technology, Hawthorn VIC 3122, Australia. His research interests include privacy preservation, differential privacy, blockchain, Internet of Things, cyber physical systems, smart grid, cognitive radio networks, and big data.  

\end{IEEEbiography}

\begin{IEEEbiography}[{\includegraphics[width=1in,height=1.25in,clip,keepaspectratio]{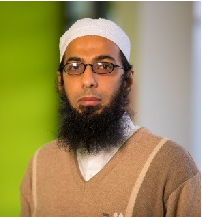}}]{Mubashir Husain Rehmani (M’14-SM’15)}

received the B.Eng. degree in computer systems engineering from Mehran University of Engineering and Technology, Jamshoro, Pakistan, in 2004, the M.S. degree from the University of Paris XI, Paris, France, in 2008, and the Ph.D. degree from the University Pierre and Marie Curie, Paris, in 2011. He is currently working as Lecturer at Munster Technological University (MTU), Ireland. He received several best paper awards. He is serving in the editorial board of several top ranked journals including NATURE Scientific Reports, IEEE Communication Surveys and Tutorials, IEEE Transactions on Green Communication and Networking and many others. He has been selected for inclusion on the annual Highly Cited Researchers™ 2020 and 2021 list from Clarivate. His performance in this context features in the top 1\% in the field of Computer Science and Cross Field.  

\end{IEEEbiography}

\begin{IEEEbiography}[{\includegraphics[width=1in,height=1.25in,clip,keepaspectratio]{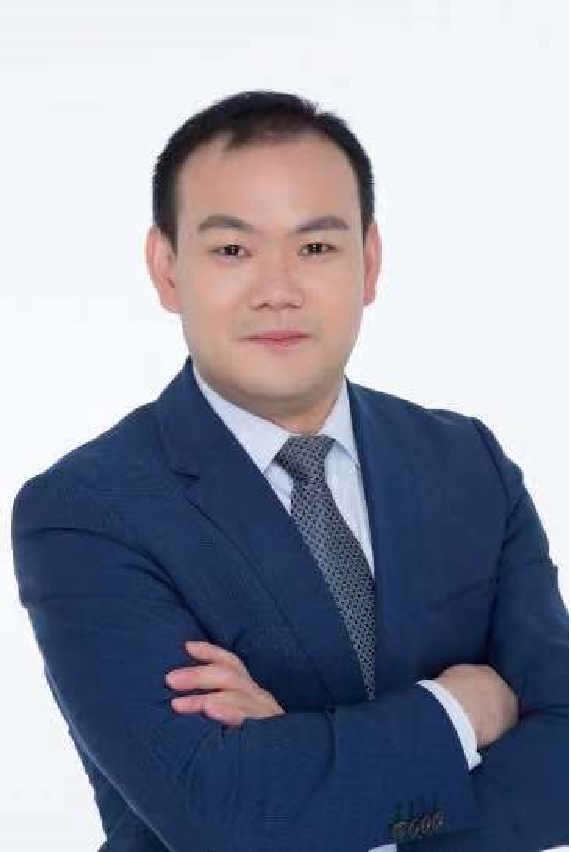}}]{Dr. Jinjun Chen}

is a Professor from Swinburne University of Technology, Australia. He is Deputy Director of Swinburne Data Science Research Institute. He holds a PhD in Information Technology from Swinburne University of Technology, Australia. His research interests include scalability, big data, data science, data systems, cloud computing, data privacy and security, health data analytics and related various research topics. His research results have been published in more than 160 papers in international journals and conferences, including various IEEE/ACM Transactions. 

\end{IEEEbiography}

\end{document}